\documentclass[letterpaper]{article}
\usepackage[T1]{fontenc}

\usepackage{geometry}
\usepackage{setspace}

\usepackage{achemso}

\usepackage{graphicx}
\usepackage{float}
\newfloat{scheme}{htbp}{los}
\floatname{scheme}{Scheme}
\floatname{chart}{Chart}
\newfloat{graph}{htbp}{loh}

\usepackage[version = 4]{mhchem} % Formulas using \ce{}
\usepackage{multirow,multicol}
\usepackage{hyperref}       % hyperlinks add hypertext capabilities
\usepackage{url}            % simple URL typesetting
\usepackage{booktabs}       % professional-quality tables
\usepackage{amsfonts}       % blackboard math symbols
\usepackage{nicefrac}       % compact symbols for 1/2, etc.
\usepackage{microtype}      % microtypography
\usepackage{xcolor}         % colors
\usepackage{graphicx}
\usepackage{amsmath,amssymb}
\usepackage[normalem]{ulem}
\usepackage{pdfpages}       %For adding PDF of SM

\usepackage{authblk}
\author[1]{Isaac L. Huidobro-Meezs*}
\author[2,3]{Jun Dai}
\author[1,4]{Rodrigo A. Vargas-Hernández*}
\affil[1]{Department of Chemistry and Chemical Biology, McMaster University, Hamilton, ON, Canada}
\affil[2]{Mila - Quebec AI Institute, Université de Montréal,  Montréal, QC, Canada}
\affil[3]{Département d’informatique et de recherche opérationnelle, Université de Montréal,  Montréal, QC, Canada}
\affil[4]{Brockhouse Institute for Materials Research, McMaster University, Hamilton, ON, Canada}

\title{Discrete Flow-Based Generative Models for Measurement Optimization in Quantum Computing}
\date{*Email: huidobri@mcmaster.ca, vargashr@mcmaster.ca}

\title{Discrete Flow-Based Generative Models for Measurement Optimization in Quantum Computing}
\date{*Email: huidobri@mcmaster.ca, vargashr@mcmaster.ca}
\begin{document}

\maketitle


\begin{abstract}
Quantum chemical simulations on quantum computers are limited by a measurement bottleneck: estimating molecular Hamiltonians to chemical accuracy requires a large number of measurements, which remain costly even on fault-tolerant devices. Hamiltonian overlapping grouping methods focus on reducing measurement counts, employing greedy initializations, such as sorted insertion (SI), leaving useful measurement/circuit trade-offs unexplored. Here, we formulate Hamiltonian grouping as a reward-driven generative search problem and introduce a Generative Flow Networks (GFlowNets)-based model that colors graph representations of molecular qubit Hamiltonians to sample non-overlapping commuting groupings. The reward function can combine measurement cost, circuit count, and compiled two-qubit-gate count, enabling multi-objective optimization without differentiable cost functions or model pretraining. Across molecular Hamiltonian benchmarks, the GFlowNets sampler finds fully commuting non-overlapping groupings with average measurement requirements 18\% lower than SI, and produces Pareto sets that expose trade-offs among shots, circuits, and two-qubit resources. When used to initialize iterative coefficient splitting (ICS), GFlowNet-generated groupings reduce post-ICS measurement estimates by up to 40\% for Jordan-Wigner-mapped fully commuting Hamiltonians relative to SI initialization. Composite rewards further identify lower two-qubit gate groupings, including cases with more than 100 fewer compiled two-qubit gates, while retaining comparable post-ICS measurement benefits. Our results show that GFlowNets provide a flexible workflow for resource-aware measurement and quantum-resource optimization in quantum chemistry, replacing single-heuristic outputs with diverse candidate groupings that can be selected according to hardware-specific priorities. Our results show that our GFlowNets' generative policy framework not only reduces measurement and two-qubit gate costs but also provides flexibility for hardware-aware adaptations via its reward function.

\end{abstract}

\section*{Keywords}
Quantum computing, Measurement optimization, GFlowNets, Measurement problem, Generative models.

% \section*{Abbreviations}

% Some journals require a list of abbreviations: these normally should be given
% immediately after the keyswords (if required).

\section{Introduction}
In the context of quantum computing for the electronic structure problem, reducing the cost of measurements remains a critical challenge for scaling up quantum algorithms. A prominent example is the variational quantum eigensolver (VQE) \cite{VQE,VQE-Rev,VQE:review:2019,VQE:review:2020,VQE:review:2020b,VQE:review:2022}, where estimating the expectation values of molecular Hamiltonians often requires a prohibitively large number of measurements to achieve chemical accuracy, constituting the measurement problem in VQE. Consequently, various strategies have been developed to reduce measurement requirements \cite{quantummeasurementquantumchemistry}.
On the algorithmic side, clever grouping of commuting operators has been shown to significantly reduce measurement overhead. Examples include splitting the Hamiltonian into simultaneously measurable fragments using (anti)commutation structure, unitary factorization, fermionic-space transformations, and symmetry considerations \cite{huggins2021efficient, oumarou2022accelerating, izmaylov2019unitary, Thomson-npj, izmaylov2020measurements, izmaylov2020JCPmeasurements, MeasFormula, Nacho2, Nacho_2023, Nachofluidfermionic}. 
These approaches generate a non-overlapping grouping of the Hamiltonian through greedy algorithms \cite{SortedInsertion,izmaylov2020JCPmeasurements} targeting fully commuting (FC) or qubit-wise commuting (QWC) groupings and use such groupings, usually from the Sorted Insertion \cite{SortedInsertion} algorithm, as initialization for overlapping grouping techniques using precomputed covariance dictionaries \cite{Thomson-npj,Choi2022,Nachofluidfermionic}. However, these optimizations don't account for costs associated with compiled circuits, such as the number of two-qubit gates and the overall circuit depth. Approaches that reduce such circuit costs rely on more flexible commutativity conditions without explicitly optimizing circuit costs by interpolating between FC and QWC, namely, $k$-commuting groups\cite{k-commutativity,Galic}.

Taken together, these measurement-reduction strategies underscore the importance of combinatorial optimization in quantum computing by balancing measurement requirements and circuit costs. Generative models and reinforcement learning (RL) are increasingly being developed to tackle problems in quantum science and related fields, including those with a strong sequential and combinatorial nature \cite{Krenn2022, AI-entanglement, zhang2023letFlows, RL4CO, NIPS2017CombOpt, CombOptNature, MultimodalLearning, TheseusDesign, InverseMolecularDesign, PRLRodrigo, Vargas-Hernandez2020, dawid2025book, ColoringPINS}. Similarly, machine learning (ML) methods are emerging as powerful tools for solving challenging combinatorial optimization problems \cite{GNNCombOpt,BENGIOCombOpt,MLCOmbOptReview}. A recent study on the elementary shortest path problem employed an unsupervised graph neural network to identify near-optimal routes, outperforming classical heuristic methods on graphs, and was capable of generalizing to much larger unseen graphs (with more than 100 nodes) \cite{ElementaryShortestPath}. These examples illustrate how AI/ML models can explore vast solution spaces, often revealing high-quality solutions or hidden patterns that conventional greedy optimization methods might overlook.

For the measurement problem, ML-based approaches have also been developed in recent years. For example, Liang et al. \cite{RL4shotAllocation} developed an RL agent that dynamically allocates measurement shots during a VQE optimization, thereby reducing the total number of measurements while still converging to the ground-state energy. 
Each of these developments, from analytic grouping to RL-based policies, helps reduce quantum measurement overhead in the current noisy intermediate-scale quantum (NISQ) era. Additionally, early evidence of efficiency gains from generative models in quantum computation is already emerging. Zou et al. \cite{GQE} demonstrated that embedding a flow-based generative model into VQE can further accelerate its convergence \cite{FlowVQEWarm}. Additionally, Nakaji et al \cite{GQE}. optimized a classical generative model to produce quantum circuits with desired properties.

Despite these advances, most current methods focus on optimizing a single objective or task (e.g., minimizing variance and identifying a single locally optimal shot distribution). Greedy strategies and RL-based algorithms often converge to a single local optimal solution, leaving other regions of the solution space unexplored. This highlights a limitation of both approaches, particularly in RL, where it is well known that agents can be trapped in a single mode of the solution space \cite{bioseq, bengio2021flow}. Restricting exploration to a single solution may overlook other viable groupings with different measurement or circuit-cost trade-offs.

Generative models offer a complementary perspective; rather than producing a single optimal solution, they aim to learn a distribution over the space of solutions. Traditionally, this class of models relies on a task-specific dataset and can be adapted to different tasks via conditional sampling schemes, making them powerful yet inherently data-dependent \cite{DeepGenerativeModelling}.
The recently proposed Generative Flow Networks (GFlowNets) overcome this limitation by providing a framework for sampling complex objects in proportion to a reward function \cite{bengio2023gflownet, bengio2021flow, zhang2023letFlows}, thereby bypassing the need for extensive training datasets. Rather than optimizing for a single objective, GFlowNets learn to generate a diverse set of high-reward solutions, effectively sampling from an unnormalized target distribution of interest, thereby allowing the inclusion of multiple figures of merit in the reward. Each sampled solution is constructed step by step, similar to an RL policy, but the learning objective ensures that the probability of generating any given solution is proportional to its reward.
This fundamentally different probabilistic approach enables GFlowNets to independently output a rich ensemble of distinct, high-quality solutions by learning a direct generative policy \cite{bengio2023gflownet}. Sampling methods such as Markov chain Monte Carlo can be considered to address this problem, but they have the disadvantage of a lengthy stochastic search of the solution space and the challenge of mode mixing \cite{MultimodalMCMC}. 

 This generative sampling approach is relevant to the measurement problem, where different valid groupings can lead to different measurement counts and circuit-cost trade-offs. By covering many modes of the solution landscape, a GFlowNets-based approach could avoid the "dominant mode" trap of conventional RL. In domains ranging from graph combinatorial problems \cite{zhang2023letFlows}, drug discovery \cite{lee2024geometricinformedgflownetsstructurebaseddrug, jain2023gflownets} to materials design\cite{MaterialsGFlow}, GFlowNets have been shown to better explore the solution space while producing high-performing candidates compared to those generated by RL or genetic algorithms. In quantum physics applications, GFlowNets have been used in simulations of open quantum systems to select a subset of Feynman paths that approximate a complex influence functional \cite{PathInt}. In this scenario, the model was able to capture the distinct high-reward modes in the problem's solution space, showcasing GFlowNets' ability to enable multimodal discovery, a desired feature for measurement optimization, as we are interested in reducing the number of measurements and circuit requirements, such as two-qubit gate counts. \\

In this work, encouraged by these developments, we recast the measurement problem as a generative search problem over valid Hamiltonian groupings and introduce GFlowNets to learn policies that sample high-quality groupings with probability proportional to a task-specific reward (e.g., an estimator inversely related to measurement requirements and circuit costs). Our construction leverages the graph-theoretic equivalence between the minimum clique cover and the coloring of the complement graph \cite{izmaylov2020JCPmeasurements}, allowing a policy to color a graph step by step while respecting validity constraints, such as disallowing the same color among neighboring nodes. Our method offers a new perspective on generating improved non-overlapping groups that can serve as initializations for overlapping methods, which are evaluated using measurement estimates and two-qubit gate requirements.
In what follows, we briefly describe the Hamiltonian grouping techniques used for molecular Hamiltonians in quantum computing in Section \ref{sec:lcu}. In Sections \ref{sec:gflownet} and \ref{sec:algo}, we introduce GFlowNets and describe the algorithm for the measurement problem, while in Section \ref{sec:arch} we discuss the architecture of our model. The results are presented in Section \ref{sec:Results}, starting with a measurement-only reward followed by a comparison with composite rewards, mixing measurement count with the number of groups produced and the number of two-qubit gates after circuit compilation.

\section{Methods}
\subsection{Grouping techniques for molecular Hamiltonians}\label{sec:lcu}
VQE techniques for the electronic structure problems are one of the most common applications of quantum algorithms in NISQ devices \cite{VQE-Rev,VQE:review:2019,VQE:review:2020,VQE:review:2020b,VQE:review:2022}. These approaches rely on measuring the expectation value of the electronic Hamiltonian, starting from the second-quantized operator projected onto a set of $N$ spin orbitals,
\begin{equation}
    H = \sum_{pq}^N h_{pq} a_p^\dagger a_q + \sum_{pqrs}^N g_{pqrs} a_p^\dagger a_q a_r^\dagger a_s, 
\end{equation}
where $a_p^\dagger\; (a_p)$ are the creation (annihilation) operators for spin orbital $p$, and $h_{pq}$, $g_{pqrs}$ are found from one- and two-body integrals. For its measurement, a fermion-to-qubit mapping is required, e.g., Jordan-Wigner (JW) \cite{JordanWigner} and Bravyi-Kitaev (BK) \cite{BK}, leading to a qubit Hamiltonian in terms of Pauli products,
\begin{equation}
       \hat{H}_q=\sum_k^{N_P} \omega_k\hat{P}_k, \ \ \ \hat{P}_k = \bigotimes_{n = 1}^{N_q} \sigma_n^{(k)},  \label{eqn:qubitH}
\end{equation}
with every Pauli product $\hat{P}_k$ being a tensor product of Pauli operators and identities for the corresponding qubit $\sigma_n \in \{\hat{x}_n, \hat{y}_n, \hat{z}_n, \hat{1}_n \}$. $N_P$ is the number of Pauli words in $\hat{H}_{q}$, and $N_q$ is the number of qubits. This number of Pauli words scales as $\mathcal{O}(N^4)$ with the number of spin orbitals considered, making direct measurement of Pauli terms impractical.

To find molecular energies, VQE performs an iterative optimization of a parameterized wavefunction, $| \psi_\theta\rangle$, as $E_\theta = \min_\theta\langle\psi_\theta|\hat{H}_q|\psi_\theta\rangle$.
Since $\hat{H}_q$ contains terms that do not commute with each other, a direct measurement of the operator is not possible. For this reason, the Hamiltonian needs to be partitioned into $N_f$ compatible fragments \cite{izmaylov2020JCPmeasurements} and measure each of them separately,
\begin{equation}
\hat{H}_q=\sum_{\alpha=1}^{N_f}\hat{H}_\alpha; \ \ \ \  E_\theta = \sum_{\alpha=1}^{N_f}\langle\psi_\theta|\hat{H}_\alpha|\psi_\theta\rangle.\label{eqn:VQE}
\end{equation}
Moreover, we require diagonalizing each fragment to make them compatible with the computational basis, transforming them into their Ising form through unitary operations \cite{quantummeasurementquantumchemistry}, incurring additional costs depending on the grouping scheme due to circuit implementation in current NISQ devices. 

Several grouping schemes exist; this work focuses on FC and QWC groupings for the VQE problem, although intermediate schemes have shown improvements in measurement allocation \cite{k-commutativity, Galic}. For FC, the requirement is that Pauli products, within a group, commute with each other $[\hat{P}_n,\hat{P}_m]=0$. QWC is a more strict condition since, for a pair of Pauli products, every single-qubit operator must commute $[\hat{\sigma}_i,\hat{\tau}_i]_{\text{QWC}}=0,\ \forall \ \sigma_i \in \hat{P}_n, \ \tau_i \in \hat{P}_m$. Both of these partitionings can be generated by constructing the corresponding commutativity graph and identifying the minimum-clique cover, with each clique in the graph representing a distinct group. This approach is equivalent to coloring the complementary graph, as is well known in graph theory and often employed in quantum computing \cite{izmaylov2020JCPmeasurements, izmaylov2020measurements}. Although efficient algorithms are known for the coloring problem \cite{Husfeldt_2015}, they do not provide any optimality guarantees for the solution. Additionally, we need to consider the cost of implementing each partitioning on current NISQ devices, i.e., FC-fragments require Clifford transformations to reach the Ising form. In contrast, QWC-fragments require single-qubit Clifford transformations, bypassing CNOT gates and their associated measurement overhead due to circuit fidelity \cite{quantummeasurementquantumchemistry}. 

Substantial efforts have been put into developing algorithms to reduce the number of measurements through grouping techniques, going from greedy algorithms \cite{SortedInsertion,Thomson-npj} to overlapping techniques, such as iterative coefficient splitting (ICS)\cite{Thomson-npj} and shared Pauli products\cite{Choi2022} which introduces "ghost" Pauli terms, and modifications of the Hamiltonian through fermionic algebra manipulations \cite{Nachofluidfermionic}, requiring iterative greedy optimizations.
Regardless, these overlapping techniques require a non-overlapping grouping as a starting point, which is frequently obtained from sorted insertion (SI) \cite{SortedInsertion}. The choice of this initialization can affect the resulting measurement estimate and circuit requirements, which we explore in the present work by generating non-overlapping groupings with GFlowNets. For a more comprehensive list of developments in this area, we refer the reader to Ref. \cite{quantummeasurementquantumchemistry}.

\subsection{GFlowNets}\label{sec:gflownet}%\vspace{-0.30cm}
GFlowNets \cite{bengio2021flow, bengio2023gflownet} are a class of generative models over a directed acyclic graph (DAG) $G=(\mathcal{S},\mathcal{A})$ of states with initial state $s_0$ and terminal set $\mathcal{S}^f$. Here, $\mathcal{S}$ denotes the set of states and $\mathcal{A}$ the set of discrete actions. The set of actions corresponds to transitions between two states $s\rightarrow s'$ with a complete trajectory $\tau\in\mathcal{T}$ defined by the sequence $\tau=(s_0\rightarrow s_1\rightarrow\dots \rightarrow s_n=s_f) $ with $s_f \in \mathcal{S}^f$. GFlowNets learn the probability of sampling a compositional object $s_f \in \mathcal{S}^f$, where the probability of reaching it is proportional to a positive reward function $R(s_f)$. 

We define a non-negative function $F: \mathcal{S} \rightarrow \mathbb{R}_{+}$, called flow, together with edge flows $F\left(s \rightarrow s^{\prime}\right) \geq 0$ for $s \rightarrow s^{\prime} \in \mathcal{A}$. The flow satisfies the conservation law at every non-terminal state $s \notin \mathcal{S}^f$,
\begin{equation}
F(s)=\sum_{s',s\to s'} F(s \to s') =\sum_{s'',s''\to s} F(s''\to s),
\label{eq:conservation}
\end{equation}
with boundary conditions at terminals $F(s_f)=R(s_f)$, and $F(s_0)=Z=\sum_{s\in\mathcal{S}^f} R(s)$. These equalities define a Markovian flow \cite{bengio2023gflownet}. The forward transition probability and the backward probability are $P_F(s' | s)={F(s \to s')}/{F(s)}$ and $P_B(s | s')={F(s \to s')}/{F(s')}$, respectively, which satisfy the detailed balance condition on the edges $F(s)P_F(s'| s)=F(s')P_B(s| s')$. 
Sampling a trajectory under $P_F$ factorizes along the DAG,
\begin{align}
    &P(\tau) = \prod_{t=1}^{n}P_F(s_{t}|s_{t-1}) \label{eqn:trajF} \\
    &P(\tau|s_n=s_f)=\prod_{t=1}^{n}P_B(s_{t-1}|s_t). \label{eqn:trajB}
\end{align}
with the terminal marginal $P_T(s_f)= R(s_f)/Z$. In GFlowNets, the training objective ensures that the flow satisfies these conservation constraints. In practice, this is achieved by parameterizing $P_F$, and optionally $P_B$, so that the conservation law holds at all internal states and the boundary matches the reward at terminals \cite{bengio2023gflownet}.
Previous work \cite{UsGFlowNets} achieved this by minimizing the violation of local consistency conditions, such as the flow matching (Eq.~\ref{eq:conservation}). Although effective, these local objectives can suffer from a slow credit assignment problem \cite{malkin2022trajectory}. The reward function $R(s_f)$ at a terminal state only provides a direct learning signal to the final transition, and this information must propagate backward through the DAG one step at a time during training. This can be inefficient for long trajectories or when the reward landscape is complex.

In this work, instead, we use the trajectory balance (TB) objective \cite{malkin2022trajectory}. TB enforces the flow-consistency condition over the entire trajectory in a single step. We have for all complete trajectories the condition,
\begin{equation}
    Z \prod_{t=1}^{n}P_F(s_{t}|s_{t-1})=R(s_f)\prod_{t=1}^{n}P_B(s_{t-1}|s_t).
    \label{eq:tb_identity}
\end{equation}
We parameterize the total flow $Z$ and the forward $P_F(s^{\prime}|s)$ and backward probabilities $P_B(s| s^{\prime})$ with a graph neural network (GNN) with parameters $\theta$. The learning objective is to minimize the squared difference loss,
\begin{equation}
    L(\tau; \theta) = \left(\log Z_\theta + \sum_{t=1}^{n} \log P_{F,\theta}(s_{t}|s_{t-1}) - \log R(s_f) - \sum_{t=1}^{n} \log P_{B,\theta}(s_{t-1}|s_t)\right)^2.
\label{eq:lossTB}
\end{equation}
Eq.~\ref{eq:lossTB} directly assigns credit for the outcome $R(s_f)$ to all transitions in the trajectory simultaneously, substantially improving the efficiency of learning in complex, high-dimensional state spaces. Minimizing the TB loss does not directly maximize the reward of a single trajectory. This loss trains the forward policy so that the induced terminal-state distribution satisfies $P_T(s_f)\propto R(s_f)$ as the reward signal is propagated back through the whole trajectory in a single step, updating $Z$, $P_F(s^{\prime}|s)$, and $P_B(s| s^{\prime})$. Therefore, terminal states with larger rewards receive larger total flow and are sampled more frequently. For a sampled trajectory, the TB loss compares the probability mass currently assigned by the model to the terminal reward; if a high-reward terminal state is underweighted, the update increases the probability of the actions leading to it, whereas overweighted low-reward trajectories are downweighted. In this way, reward information is propagated through the full trajectory, improving credit assignment and progressively biasing sampling toward higher-reward states while still preserving diversity among solutions \cite{bengio2023gflownet,malkin2022trajectory}. 

GFlowNets thus offer a principled method for sampling from a distribution proportional to a reward function \( R(s) \), or equivalently from an energy function \( E(s):= -\log R(s) \) \cite{bengio2023gflownet}, allowing custom rewards tailored to specific problems. Unlike most generative models, the training protocol occurs concurrently with the sampling stage. Furthermore, GFlowNets-based approaches are particularly effective in problems exhibiting a compositional structure, where sequential generation is natural, the reward function is non-negative and easy to evaluate, and the target distribution is highly multimodal \cite{jain2023gflownets}.

\subsection{Color assignment with GFlowNets}\label{sec:algo}
For the measurement problem, the algorithm consists of sequentially coloring the graph that represents the qubit Hamiltonian. The goal is to color the complement of the commutativity graph of the system, where each node represents a Pauli word in the Hamiltonian, and the node color indicates which group it belongs to. 

Starting from the qubit Hamiltonian $\hat{H}_q=\sum_{k=1}^{N_P}\omega_k\hat{P}_k$ of Eq. \ref{eqn:qubitH}, we define a commutativity graph $G_{\mathrm{comm}}=(V,E_{\mathrm{comm}})$ with one vertex $v_k\in V$ for each Pauli word $\hat{P}_k$. The order of the vertices is taken directly from the  OpenFermion \cite{openfermion} Qubit Hamiltonian object. Alternatively, one can order the nodes by the weights of the Pauli operators, $|\omega_k|$, as discussed later. For FC grouping, $(v_i,v_j)\in E_{\mathrm{comm}}$ when $[\hat{P}_i,\hat{P}_j]=0$; for QWC grouping, the edge is present when the two Pauli words commute qubit by qubit. A compatible measurement group is therefore a clique of $G_{\mathrm{comm}}$, and a non-overlapping grouping is a clique cover. Clique cover problems are equivalent to the coloring of the complement graph \cite{Husfeldt_2015,izmaylov2020JCPmeasurements}. The proposed algorithm colors the complement of the commutativity graph, which we will call the conflict graph, defined as $G_{\mathrm{conf}}=\overline{G}_{\mathrm{comm}}$. An edge of this conflict graph joins two Pauli words that cannot be placed in the same measurement group. A proper coloring $c:V\rightarrow\{1,\ldots,K\}$ satisfies $c_i\neq c_j$ for every $(v_i,v_j)\in E_{\mathrm{conf}}$, and each color class is exactly one FC or QWC fragment.

The GFlowNets protocol takes the uncolored conflict graph of the system as initial input. The node ordering is the same as the qubit Hamiltonian object from OpenFermion \cite{openfermion}; however, we also consider in Sec. \ref{sec:Node} node orderings based on Pauli word weights $|\omega_k|$. This state is given to a GNN, as described in Section \ref{sec:arch}, which will estimate the forward flow of assigning a color to the next node, $P_F$, alongside the associated backward policy $P_B$. The color assignment on each node, with available colors, represents the set of actions for GFlowNets and is masked according to validity constraints; i.e., a graph is invalid if two connected nodes have the same color. The step in the trajectory determines which node to color. To reduce the size of the sampling space, we set an upper bound on the number of colors according to a greedy coloring algorithm, which tries to color the graph using as few colors as possible, following a random sequential strategy, as implemented in NetworkX \cite{NetworkX}; this bound is then increased by two to improve the flexibility of the solution space and implemented in the mask function as a color cap $K_{\max}$. Because that greedy coloring is itself valid, its number of colors is a feasible upper bound on the chromatic number. At a partial state, the mask removes any color in $\{1,\ldots,K_{\max}\}$ that is already assigned to a colored neighbor of the active node.

\begin{figure}[t!]
    \centering
    \includegraphics[width=\linewidth]{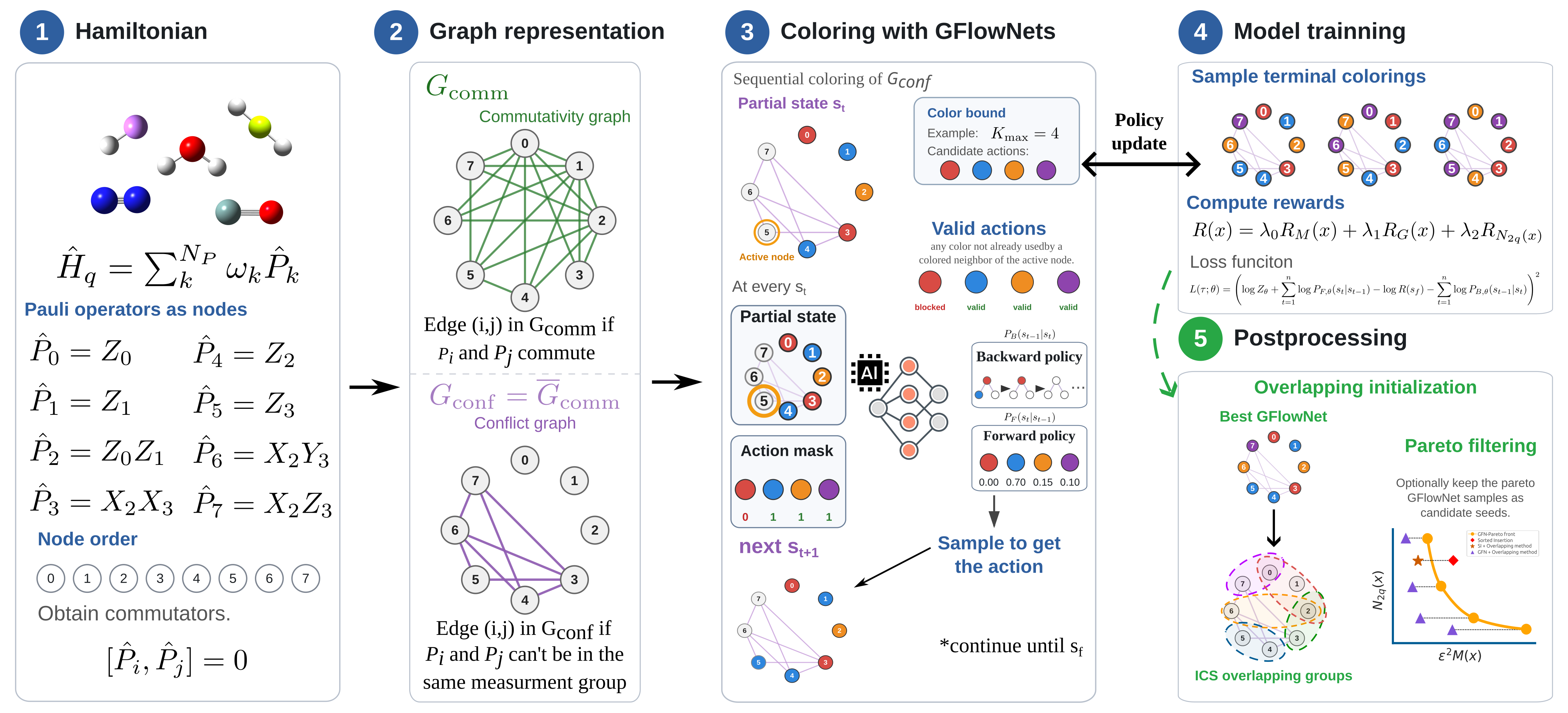}
    \caption{Diagram of the GFlowNets sampling process. Given a qubit Hamiltonian, the algorithm constructs the commutativity graph and its complement, which we label the conflict graph. The initial state is an uncolored graph, for which different trajectories are generated. The set of actions corresponds to the color assignment to each node. At each step of object construction, the GFlowNets policy provides the forward and backward probabilities, which are parametrized by a GNN. Sampling from the forward policy gives the next action in the object construction. States that lead to invalid graphs (two neighbors with the same color) are masked from the set of actions. The reward function is computed for fully colored graphs; if a coloring is invalid, the reward function goes to zero. The model is updated after \texttt{n\_update} full trajectories by minimizing the trajectory balance loss function. When sampling is complete, the resulting graphs can be postprocessed to initialize overlapping grouping methods or to select a grouping based on the Pareto front. }
    \label{fig:GFlow}
\end{figure}

Our proposed algorithm continues until a fully colored graph is produced, representing a terminal state $s_f$ corresponding to a non-overlapping grouping of the Hamiltonian, for which a reward function $R(s_f)$ is computed. The loss function is generated for the trajectory, and the model is updated after \texttt{n\_update} full trajectories. The algorithm stops when the preset number of total samples is generated. The approximate computational complexity of the algorithm is $\sim\mathcal{O}\left(S(N_P^2+E)\right)$ where $S$ is the number of samples and $E$ the number of edges on the graph, which for sparse graphs becomes $\sim\mathcal{O}\left(SN_P^2\right)$.  See Fig.~\ref{fig:GFlow} for a visualization of the proposed algorithm. \\

To sample useful and valid groupings with GFlowNets, the reward function must be carefully designed. We define the total reward $R(x)$ as a weighted sum of three components: i) the number of measurements required to reach a target accuracy $R_M$, ii) a regularization term penalizing excessive groupings (colors) $R_G$, and iii) the total number of two-qubit gates for the compiled circuits for a given grouping
\begin{equation}
R(x) = \lambda_{0} R_M(x) + \lambda_{1} R_G(x) + \lambda_2R_{N_{2q}(x)}, \label{eqn:reward}
\end{equation}
% where $\lambda_{0}$ and $\lambda_{1}$ are hyperparameters controlling the relative importance of each term. 
where $\lambda_i$s are the set of hyperparameters that control each term's relevancy in $R(x)$.
The measurement term is given by,
\begin{eqnarray}
R_M(x) &=& \frac{1}{\varepsilon^2M(x)}, \label{eqn:reward_m} \\
\varepsilon^2M(x) &=&  \left( \sum_{\alpha=1}^{N_f} \sqrt{\mathrm{Var}(\hat{H}_\alpha)} \right)^2, \label{eqn:meas}
\end{eqnarray}
where $M(x)$ is the total number of measurements needed to achieve a target accuracy $\varepsilon$ \cite{MeasFormula, Thomson-npj}, and $\mathrm{Var}(\hat{H}_\alpha)$ is the variance of the fragment $\alpha$.
In this work, we compute variances from the exact full configuration interaction (FCI) wavefunction. Setting $\varepsilon=1$ mHa, below chemical accuracy ($< 1.6$ mHa), allows to read the $\varepsilon^2M$ values as the number of measurements in millions to reach a $\varepsilon=1$ mHa error. Classically efficient wavefunctions can also be used to estimate the variances, introducing less than 9\% error for systems similar to those studied here \cite{Nachofluidfermionic}; we include results from training with Configuration Interaction Singles and Doubles (CISD) in the Supplemental Material (SM) to exemplify this, showing differences of $\varepsilon^2M(x)$ consistent with the cited results. Other proxy wavefunctions from pre-trained models like coupled-cluster-inspired constructions \cite{Mole} or approximations based on the 1- and 2-body reduced density matrices \cite{ML1RDM,b2026ML2RDM} can be employed to estimate the variances without the need for an electronic structure calculation. We report the results of training with FCI, since the resulting models can serve as a warm-up stage for future work on transferable generative models that produce non-overlapping groupings for n-qubit systems, as these Hamiltonians are subgraphs of the Pauli group on $n$ qubits. 
The second term in the reward,
\begin{equation}
R_G(x) = N_P - N_G(x), \label{eqn:reward_g}
\end{equation}
regularizes the number of groups ($N_G$), accounting for the number of quantum circuits required, and is given by $N_G=\text{max}\_\text{color}(x)$ where $\text{max}\_\text{color}(x)$ is the maximum number of distinct colors found in the graph.
The third term,
\begin{equation}
    R_{N_{2q}}(x)=\frac{1}{N_{2q}(x)}, \label{eqn:reward_2q}
\end{equation}
takes into account the total number of two-qubit gates ($N_{2q}$) across the measurement circuits for FC groups. Measurement circuit compilation is performed with Tequila\cite{tequila}, in an all-to-all architecture, using the Patel, Markov, and Hayes algorithm \cite{PatelMarkovHayes}. We explore only this compilation strategy; however, other strategies tailored to specific device architectures \cite{SynthesisCircuitsDiffusionModel,SynthesisAlphaTensor,SynthesisQMAP,SynthesisGraphs,SynthesisIBMQX,SynthesisSU4,SynthesisFieldNeutralAtoms} can be incorporated into the reward. It is essential to note that the reward function can include any quantity derived from the terminal object, and it is important to note that it is not required to be differentiable.
This allows the sampling distribution to be biased toward candidate groupings with fewer quantum circuits or lower two-qubit counts rather than focusing solely on minimizing the number of measurements. Additional optimization objectives can be incorporated into the reward, for example, fidelity overheads \cite{ZachFidOverhead} arising from circuit implementation or restrictions on hardware-efficient measurement circuits \cite{Hardware-efficientMeasurements,MillerHardwareTailored}. It is important to emphasize that the reward function can be freely modified to meet the user's needs by adjusting the $\lambda$ parameters or including additional functions as needed. Moreover, the non-overlapping groups generated by GFlowNets can be employed as initializations for overlapping methods, as we show for ICS.
\begin{figure}[t!]
    \centering
    \includegraphics[width=0.5\linewidth]{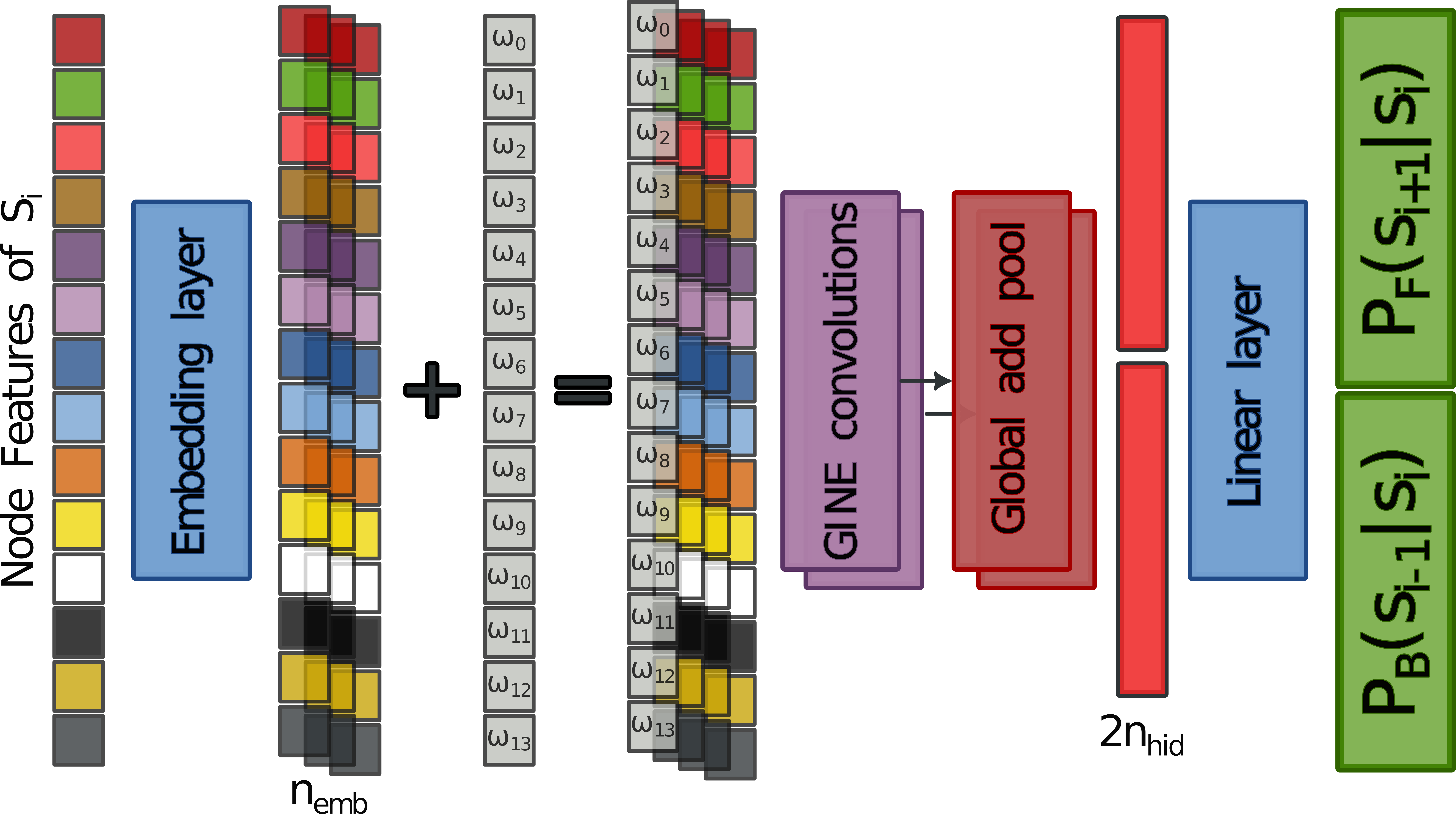}
    \caption{Diagram for the $\texttt{GINE}_w$ model architecture. The values $n_{emb}$ and $n_{hid}$ are equal to \texttt{emb\_d} and \texttt{hidden\_d}. After passing the node colors through an embedding layer, the representation is augmented by the coefficients of Eq. \ref{eqn:qubitH}. For the \texttt{GINE} model, we simply discard the coefficients vector, added after the embedding.}
    \label{fig:GINE}
\end{figure}

\subsection{Flow function architecture}\label{sec:arch}
In the proposed GFlowNets-based framework for measurement optimization, the forward and backward policies are computed by the global add pooling operation after two graph convolution layers. 
Our GNN architecture is based on the graph isomorphism network~\cite{GIN:2019}, modified to incorporate edge features (GINE)~\cite{GINE}. 
The node update rule is given by,
\begin{equation}
    x_i' = h_\theta\left( x_i + \sum_{j \in \mathcal{N}_i} \text{ReLU}(x_j + e_{j,i}) \right),\label{eq:GINE}
\end{equation}
where $x_i$ are the input node features, $x_i'$ are the updated node features, corresponding here to the forward and backward policies for object generation, $\mathcal{N}_i$ is the set of neighbors of node $i$, $e_{j,i}$ are the edge features, and $h_\theta$ is a multilayer perceptron (MLP).

The GNN input represents a state in which each node corresponds to a term of the Hamiltonian (Pauli words), and its feature encodes the associated color or group. All edges are initialized with a feature value of $-1$ to encode anticorrelation, since neighbors can't share the same color in the complementary commutativity graph. 
The architecture begins with an embedding layer with dimension \texttt{emb\_d}, followed by two GINE convolutional layers, each containing a two-layer MLP with dimension \texttt{hidden\_d} and ReLU activations. 
The logits for the sequential coloring probabilities are produced by a two-layer MLP with the same \texttt{hidden\_d} size. These logits are then used to sample from a categorical distribution with a fixed seed for all simulations, outputting the action for each step in the state construction. We denote this model as $\texttt{GINE}$ throughout the manuscript. A variant architecture, denoted $\texttt{GINE}_w$, follows the same structure but augments the node embedding with the weights of the Pauli words from Eq. \ref{eqn:qubitH} as an extra feature dimension. To visualize this architecture, see Fig. \ref{fig:GINE}. We found that for most systems, this architecture outperforms the initial \texttt{GINE} implementation.

\section{Results and Discussion}\label{sec:Results}
We split our results into two sections: Section \ref{sec:MO}, which examines a measurement-only reward, and Section \ref{sec:Composite}, which analyzes the effect of different hyperparameter values on the reward (Eq.~\ref{eqn:reward}). All molecular Hamiltonians were generated using the Tequila package \cite{tequila}, employing the PySCF \cite{PySCF,PySCF2} backend, and the STO-3G basis set \cite{STO-3G}. The interatomic distance is set to 1 {\AA} for all molecules, except for \ce{SiO}, which used a distance of 1.5 {\AA}, close to the equilibrium geometry. For \ce{H2O}, an angle of $107.6^\circ$ was used. These geometries are intended only as a benchmark for testing our algorithm, motivated by previous studies  \cite{Thomson-npj,Nachofluidfermionic,Choi2022}. For model optimization, we use $\texttt{n\_update}=10$ samples per optimization step and a learning rate of $3\times10^{-4}$ with the Adam optimizer \cite{adam}. We compare our results with the recursive largest-first (RLF) and sorted insertion (SI) algorithms \cite{SortedInsertion}, which are also implemented in the Tequila suite. These algorithms generate non-overlapping groupings without requiring covariance-based minimization, making them appropriate baselines for the non-overlapping grouping stage considered here. We also consider a greedy algorithm that uses covariance information to produce non-overlapping groupings. The Greedy grouping algorithm constructs measurement groups using Pauli-term variances and covariances evaluated with respect to a reference wavefunction, iteratively selecting the compatible term--group assignment that yields the lowest estimated measurement cost. New groups are created only when a term cannot be assigned to any existing compatible group. For details, see SM. SI can be shown to minimize measurement costs under the assumptions of zero covariance and unit variances, as discussed later.
We use these methods as a baseline for the proposed generative model. For \ce{H2}, \ce{H4}, \ce{LiH}, \ce{BeH2} we employed $5,000$  total samples, for \ce{H2O} total $1,000$, and for \ce{N2} and \ce{SiO}, $500$ total samples. The reduced number of samples follows from the computational scaling discussed in SM Section 1, as for a fixed wall-clock budget the algorithm would, to first order, permit a sample count that decreases as $S\propto 1/(N_P^2+E)$. Typical color bounds employed for each system during sampling can be found in the SM Section 3.
For larger systems, a sample reduction was necessary due to the lengthy trajectories required to generate a valid graph.
We observed no GPU speedup; therefore, all simulations were performed on AMD EPYC 9655 (Zen 5) processors running at 2.6 GHz with 384 MB cache.

\subsection{Measurement-only reward}\label{sec:MO}

To assess the proposed GFlowNets-based grouping protocol under a measurement-only reward, we first employ a reward tailored solely for this task. For all experiments in this subsection, we set the values $\lambda_0=10^3$, and $\lambda_1 = \lambda_2=0$ in Eq.~\ref{eqn:reward}, resulting in a measurement-only reward. We consider the two architectures discussed in Section~\ref{sec:arch} and find that, in most cases, $\texttt{GINE}_w$ outperforms $\texttt{GINE}$. The summary of our results for the FC grouping can be found in Table~\ref{tab:ExactVarFC}. No substantial differences were observed between the JW and BK mappings for the FC grouping with the generated Hamiltonians, so we present only the JW results here. This behavior is consistent with the fact that the JW and BK encodings are related by an invertible binary linear transformation of the computational-basis representation \cite{BK2,chiew2024ternarytreetransformationsequivalent} that can be implemented by Clifford operations. Full commutation is preserved under Clifford conjugation, so the structure of the conflict graph for FC groupings is therefore not expected to depend on the chosen mapping, as can be seen in Tables SM-4 and SM-5. 

{\renewcommand{\arraystretch}{1.3}
\setlength{\tabcolsep}{2pt}
\begin{table}[t!]
    \centering
    \begin{tabular}{|c||c|c|c|c|c||c|c|} \hline
        System [$N_P$] & \texttt{GINE}  & $\texttt{GINE}_w$ & Greedy  & RLF & SI & SI-ICS & GFN-ICS \\ \hline
        \ce{H2} [14] & \textbf{0.136 (2)} & \textbf{0.136 (2)} & \textbf{0.136 (2)}&  \textbf{0.136 (2)} & \textbf{0.136 (2)}  & \textbf{0.136 (2)}& \textbf{0.136 (2)}\\ \hline
        \ce{H4} [184]& \textbf{0.759 (11)} & 0.805 (11)& 1.851 (12)& 1.424 (8)& 1.015 (9) & 0.705 (9) & \textbf{0.434 (11)}\\ \hline
        \ce{LiH} [275]& 0.262 (20) & \textbf{0.224 (18)}& 0.496 (22) & 1.014 (17) & 0.276 (20) &0.122 (20) & \textbf{0.096 (18)} \\ \hline
        \ce{BeH2} [326]&  0.643 (13)& \textbf{0.601 (21)}& 0.952 (17)& 1.736 (16) & 0.614 (18)  & 0.301 (18) & \textbf{0.234 (21)} \\ \hline
        \ce{H2O} [550]&  4.056 (33) & 4.002 (29) & 5.675 (31) & 5.521 (24)& \textbf{2.783 (30)} & 1.005 (30)& \textbf{0.931 (29)}\\ \hline
        \ce{LiH}$^*$ [630]& 0.834 (36) & \textbf{0.769 (32)}& 1.135 (38) &1.018 (26) & 0.882 (42) &\textbf{0.233 (42)}& 0.255 (32) \\ \hline
        \ce{N2} [824]& 3.299 (25) & \textbf{3.080 (22)}& 7.965 (31) &5.895 (19)  & 3.782 (26) & 1.612 (26)& \textbf{1.514 (22)} \\ \hline
        \ce{SiO} [1616]& -- & \textbf{15.123 (66)} & 26.530 (69) & 17.810 (47) & 17.829 (55) & \textbf{1.992 (55)} & 2.533 (66) \\ \hline
    \end{tabular}
    \caption{Comparison of $\varepsilon^2M(x)$ values in the FC grouping with GFlowNets protocol using the $\texttt{GINE}_w$ and $\texttt{GINE}$ models. These values are equivalent to the number of measurements (in millions) required to achieve 1 mHa accuracy. Number of groups $N_G(x)$ reported in parentheses. Results in bold signal the solutions with the lowest number of $\varepsilon^2M(x)$ for both the non-overlapping and overlapping settings. The \ce{LiH}$^*$ Hamiltonian was taken from Ref.~\cite{Thomson-npj} , which employs the BK mapping, for which 500 total samples were employed. The SI-ICS overlapping method uses as a starting point the groups from Sorted Insertion, while GFN-ICS uses the groupings from GFlowNets with the $\texttt{GINE}_w$ model.}
    \label{tab:ExactVarFC}
\end{table}}

As shown in Table \ref{tab:ExactVarFC}, the GFlowNets-based model can find solutions with lower measurement estimates than the RLF and SI algorithms for all molecules, except for \ce{H2O}. We will provide an in-depth analysis for this in Section \ref{sec:Node}. Both \texttt{GINE} and $\texttt{GINE}_w$ models outperform the SI algorithm; however, the more compact \texttt{GINE} model achieved better results for the \ce{H4} molecule. For $\texttt{GINE}_w$, we found that an architecture with $\texttt{emb\_d}=2$ and $\texttt{hidden\_d}=64$ performed the best for most of the molecules. Yet, for \ce{N2} and \ce{SiO},  $\texttt{hidden\_d}=8$ and $\texttt{hidden\_d}=24$, respectively, with $\texttt{emb\_d}=2$  were sufficient. Higher values of \texttt{hidden\_d} did not improve results and increased the model's memory requirements due to the high node count of this graph. For \ce{SiO} we decided to employ only the $\texttt{GINE}_w$ as it consistently performed better; thus, the entry for \texttt{GINE} is omitted in Table \ref{tab:ExactVarFC}. Our testing with low-dimensional embeddings is consistent with previous research on GNN performance \cite{LowDEmb,node2vec}. An initial implementation of this GFlowNets algorithm employed a simple 2-layer MLP to output the coloring probabilities from a vector containing the color of each node~\cite{UsGFlowNets}. As shown in Table SM-6, this naive approach yielded abysmal results. For most molecules, GFlowNets with a 2-layer MLP were unable to find better grouping schemes than RLF, highlighting the gains from leveraging the graph structure of the problem through a GNN. Additional results for other architectures are detailed in Tables SM-7, SM-8, and SM-9 in the SM.\\

\begin{figure}[t!]
    \centering
    \includegraphics[width=0.85\linewidth]{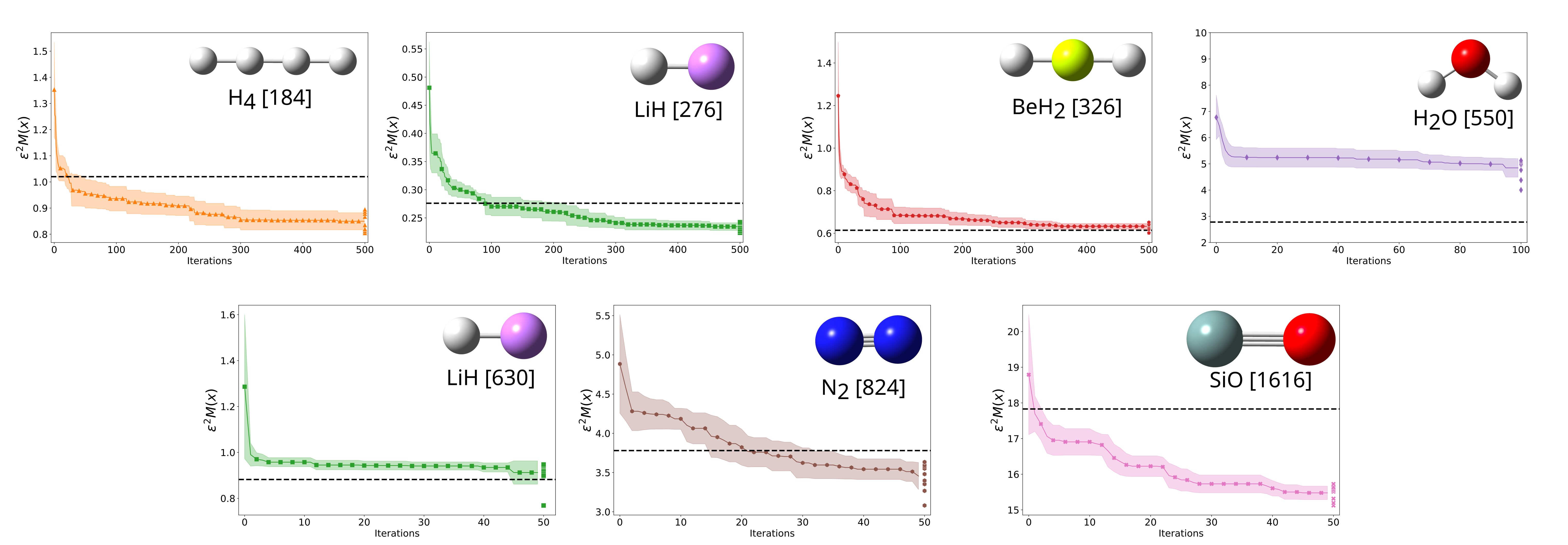}
    \caption{Average of the top 10 samples, at each iteration, according to their reward over the sampling procedure. $N_P$ shown within the brackets. The model employed is the $\texttt{GINE}_w$. The points at the end of the graph are the overall best 10 samples. The Hamiltonian for LiH with $N_P=630$ was taken from Ref.~\cite{Thomson-npj} . Measurements obtained via sorted insertion are shown as a black dotted line in each graph.}
    \label{fig:CumulativeBest}
\end{figure}

To observe the behavior of the algorithm over the sampling stage, we plot in Fig. \ref{fig:CumulativeBest} the top 10 samples, according to their measurement value $\varepsilon^2M(x)$, throughout the iterations for the $\texttt{GINE}_w$ model. Each iteration contains 10 samples, after which the model is updated by minimizing the trajectory balance loss, Eq. \ref{eq:lossTB}. The shaded area represents the standard deviation of the 10 best samples collected until that iteration. In general, we observe an initial decrease in the average number of measurements, followed by sequential improvements as the algorithm finds better solutions. This behavior reflects the reward-biased sampling of valid terminal groupings. As the algorithm discovers new high-performing solutions, the probabilities are adjusted based on the reward. However, they are not set to 0 unless the solution is invalid, allowing for diverse exploration of the solution space.

\begin{figure}[h!]
    \centering
    \includegraphics[width=0.35\linewidth]{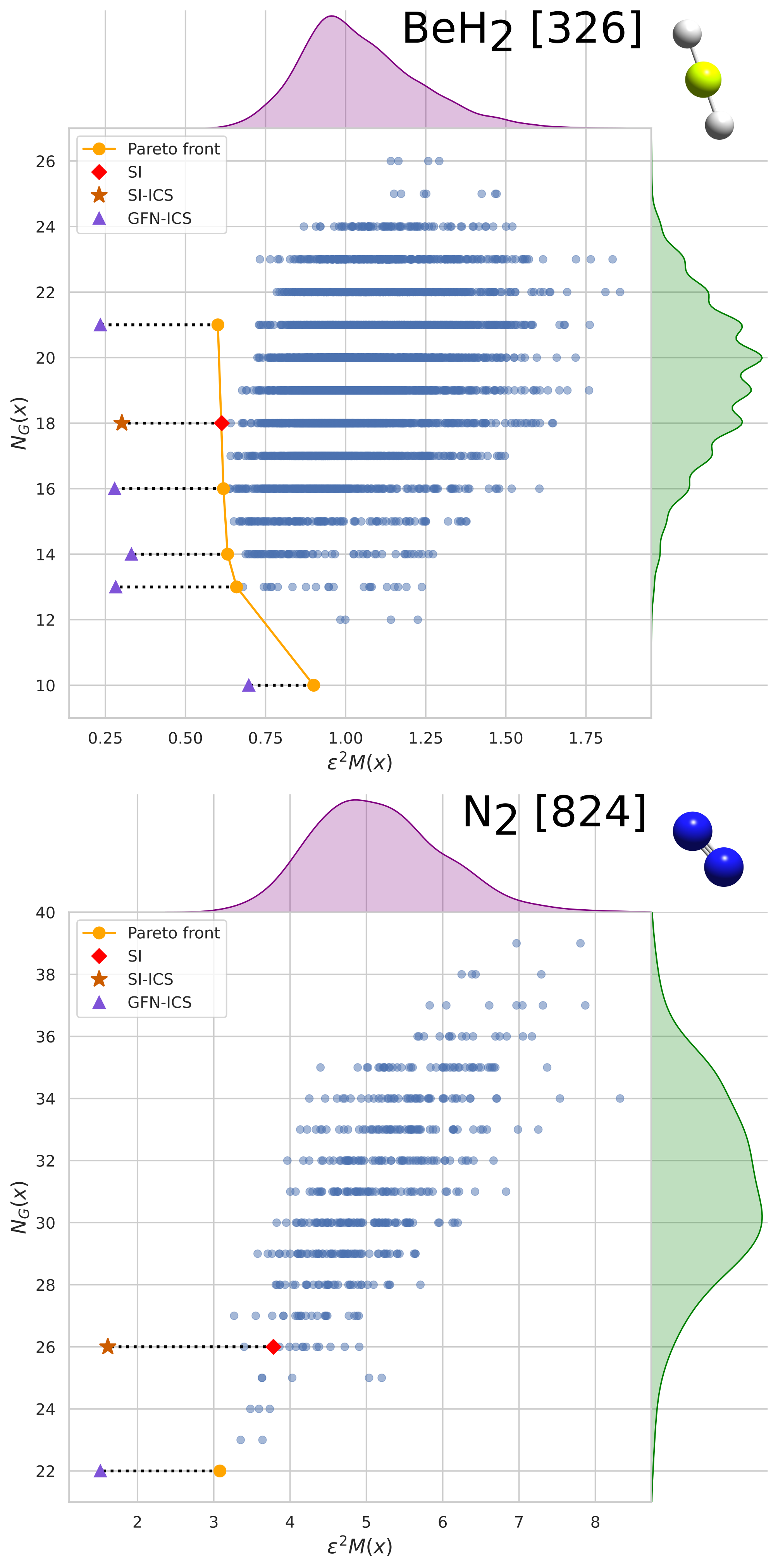}
    \caption{Pareto front for \ce{BeH2} and \ce{N2} with JW mapping and FC grouping. \ce{BeH2} and \ce{N2} panels contain 5,000 and 500, respectively. The histograms in the margins show the marginal distributions of $\varepsilon^2 M(x)$ and $N_G(x)$ across the total sampled circuits for each system. ICS results using the groupings present in the Pareto front as initialization are shown as GFN-ICS.}
    \label{fig:Pareto}
\end{figure}

Fig. \ref{fig:Pareto} shows the Pareto front\footnote{The \emph{Pareto front} is defined as $\{x \in \mathcal{X} | \nexists\, y \in \mathcal{X} : f(y) \leq f(x) \wedge f(y) \neq f(x)\}$.} from the collected samples by GFlowNets, and also the smoothed distributions of the sampled measurement and number of groups requirements. Smoothed distributions were generated using kernel density estimation (KDE) with a Gaussian kernel and a 0.8 smoothing bandwidth factor, employing Scott's rule \cite{scott1992multivariate}. For \ce{BeH2}, GFlowNets found solutions with up to 8 fewer groups than SI, and showcases similar performing results with 4 fewer groups, with $(\varepsilon^2M, N_G)=(0.632, 14)$ representing only a 2.9\% measurement increase with respect to SI. For \ce{N2}, remarkably, despite the low number of samples, our method encounters better solutions before reaching 20 iterations (a total of 200 groupings), see Fig. \ref{fig:CumulativeBest}, and represents a significant improvement in both number of groups and measurements; see also Fig.~\ref{fig:Pareto}. In Figs. SM-1 and SM-2, we present the histograms and the generated Pareto fronts for the remaining molecules using the \texttt{GINE} and \texttt{GINE}$_w$ models, respectively, for the FC grouping. \ce{H4} presents as well interesting solutions as a result of the sampling nature of GFlowNets, finding a grouping remarkably close to the one with the lowest measurement count for the \texttt{GINE}$_w$ model ($\varepsilon^2M=$0.805 and 11 groups vs 0.812, 9 groups), producing an almost vertical line in the Pareto front for that experiment as can be seen in Fig. SM-2.\\

One potential application of our algorithm is to provide a better initialization for overlapping methods. We adapted our graph generation to include compatibility functions that enable the generated groupings to be used with the \emph{Tequila} library \cite{tequila}, which includes the ICS method. Using Tequila's code, we initialize the ICS method with the GFlowNets non-overlapping groupings generated with the \texttt{GINE}$_w$ model. This strategy will be abbreviated as GFN-ICS from here on forward, while initialization with SI will be SI-ICS. Table. \ref{tab:ExactVarFC} shows the measurement results for such initialization. The results show that GFN-ICS consistently obtained lower measurement counts than SI-ICS for the JW mapped Hamiltonians, taking the loss by a small margin for the \ce{LiH}$^*$ Hamiltonian, which employs the BK mapping, for which only 500 total samples were employed. In this case, despite having a larger $\varepsilon^2M$ value by 0.022, the GFN-ICS solution requires 10 groups less, which translates to fewer quantum circuits to run. For the JW mapped molecular Hamiltonians studied here, GFN-ICS got a reduction ranging from 6\% to 38\%, getting an average reduction of 11.4\% with respect to the SI-ICS baseline. This comparison isolates the effect of replacing the SI initialization within the same ICS workflow. Moreover, Fig. \ref{fig:Pareto} shows that, for \ce{BeH2} and \ce{N2}, the Pareto-front candidates have lower or comparable estimated measurement counts while requiring fewer quantum circuits. This trend is followed for the other molecules studied here, as shown in the SM Fig. SM-2.

\begin{figure}[t!]
    \centering
    \includegraphics[width=0.40\linewidth]{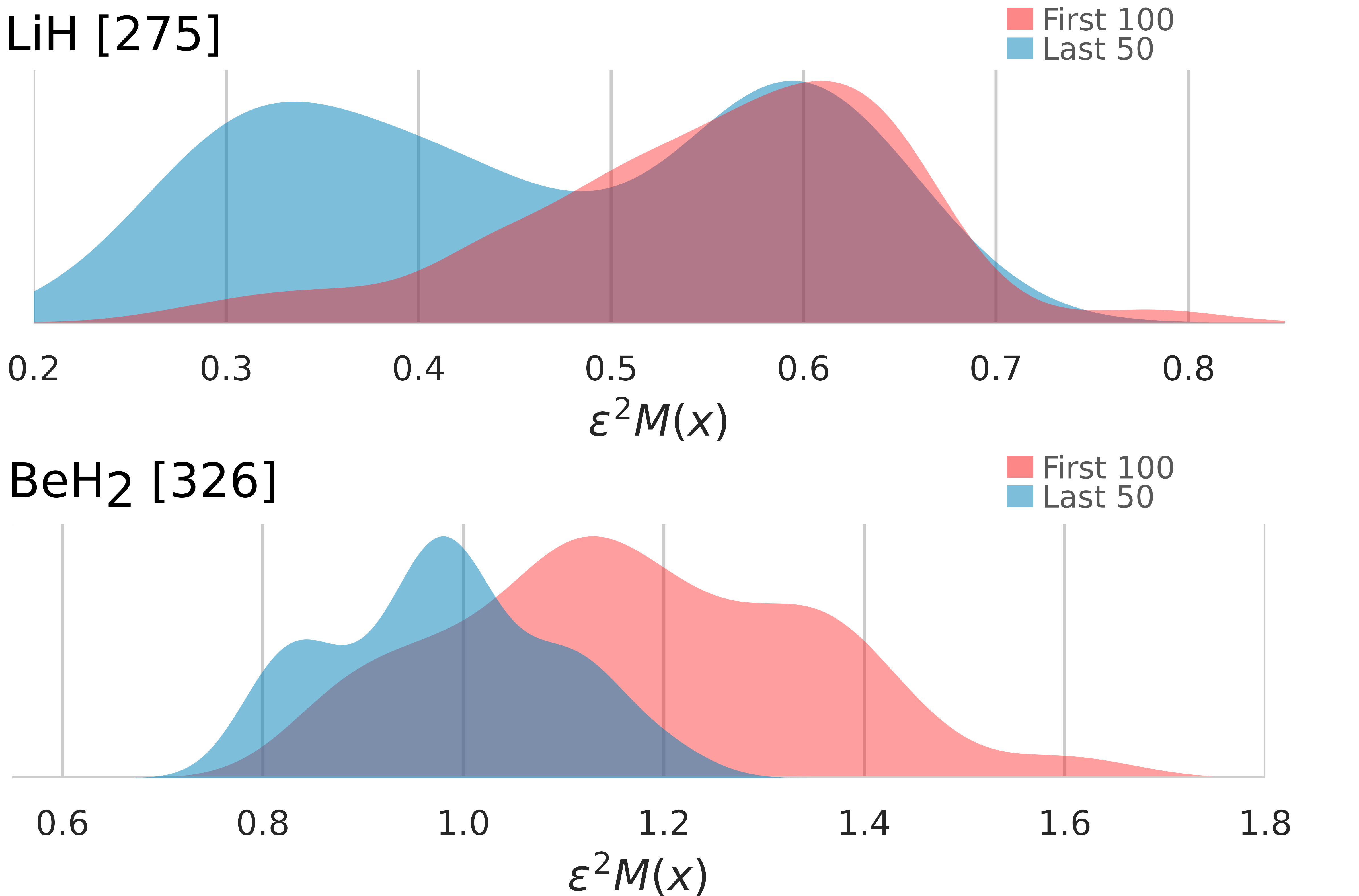}
    \caption{Smoothed histogram distributions of the $\varepsilon^2 M(x)$ values for \ce{LiH} and \ce{BeH2}. Each panel shows the distribution from the first 100 and the last 50 sampled groupings (of 5,000 total) produced by GFlowNets, illustrating how the samples evolve during optimization.}
    \label{fig:KDEevolve}
\end{figure}

To further emphasize the active learning of our GFlowNets algorithm, we show in Fig. \ref{fig:KDEevolve} the histograms, smoothed with the KDE, in terms of the number of measurements, Eq. \ref{eqn:meas}, for the first 100 and last 50 samples for \ce{LiH} and \ce{BeH2}. In both cases, we observe a displacement of sample density towards lower measurement counts, indicating a shift in the sampled groupings towards lower measurement counts.

For QWC grouping, we consider both JW and BK mappings since the SI algorithm yields markedly different outcomes. This is expected as qubit-wise commutation is not invariant under an entangling Clifford transformation, because it depends on the single-qubit Pauli operator present at each position. Consequently, the JW and BK QWC conflict graphs can differ substantially. As shown in Table SM-3, the GFlowNets algorithm does not outperform the baseline for \ce{BeH2} and \ce{H2O} with the employed GNN architectures. For the \texttt{GINE}$_w$ model, we used $\texttt{emb\_d}=2$ and $\texttt{hidden\_d}=64$ for \ce{H4}, \ce{LiH}, \ce{BeH2}, and \ce{H2O}. In the case of \ce{H2O} with the BK mapping, we reduced the hidden dimension to $8$ due to memory constraints arising from its densely connected graph (see Table SM-5). We attribute the limited performance to architectural restrictions that hinder the generation of valid low-measurement groupings as the number of groups increases. This issue relates to the well-known neighbor explosion problem in GNNs \cite{xue2025haste,pmlr-NeighborExpFeatMom,LargeScaleGraphsNeighbExp}. Scaling to larger architectures is impractical due to memory demands, rendering the current version of the algorithm inefficient.

As shown in Tables SM-4 and SM-5, for the complement of the commutativity graph, the average degree of each node increases drastically when going from FC to QWC, thus increasing the number of message-passing operations in Eq. \ref{eq:GINE} and the difficulty of this task in conjunction with higher memory requirements. Throughout our experiments, we observed that in the QWC grouping, most of the sampling effort was spent learning to generate valid graphs rather than sampling groupings with lower estimated measurement counts. In contrast to the FC experiments, where 100\% of graphs were valid in most cases and only a few cases produced around 1\% invalid graphs, QWC sampling efficiency drops drastically. Employing the \texttt{GINE}$_w$ model, for JW mappings, the proportion of valid graphs for \ce{BeH2} drops to 85\%, \ce{H2O} to 54\%, and \ce{H4} all the way to 15\%. For BK, the problem is starker, with none of the molecules achieving more than 30\% valid graphs. This reduction in efficiency is evident in the histograms shown in Figs. SM-3 and SM-4, where we also observe a similar behavior for the \texttt{GINE} model.

Since the masking function employed has an upper bound based on the number of colors, we relaxed this condition to the maximum node degree for each graph in \ce{BeH2} and \ce{H2O} for JW and BK mappings and \ce{H4} for BK mapping, given that these were the cases in which GFlowNets couldn't find better solutions. This ensures 100\% valid samples but risks finding solutions close to the trivial one, since the maximum degree is close to the number of nodes in the graph, \ce{H4}$_{\text{,BK}}$: (Nodes=184, $\max$ degree=183), \ce{BeH2}$_{\text{,JW}}$:(326, 289), \ce{BeH2}$_{\text{,BK}}$:(326, 310), \ce{H2O}$_{\text{JW}}$:(550, 513), \ce{H2O}$_{\text{BK}}$:(550, 537). We also ran the simulations with around half the maximum degree as the upper bound on the number of colors; however, both approaches, while producing valid states, were unable to find better solutions. 

\begin{figure}[h!]
    \centering
    \includegraphics[width=0.35\linewidth]{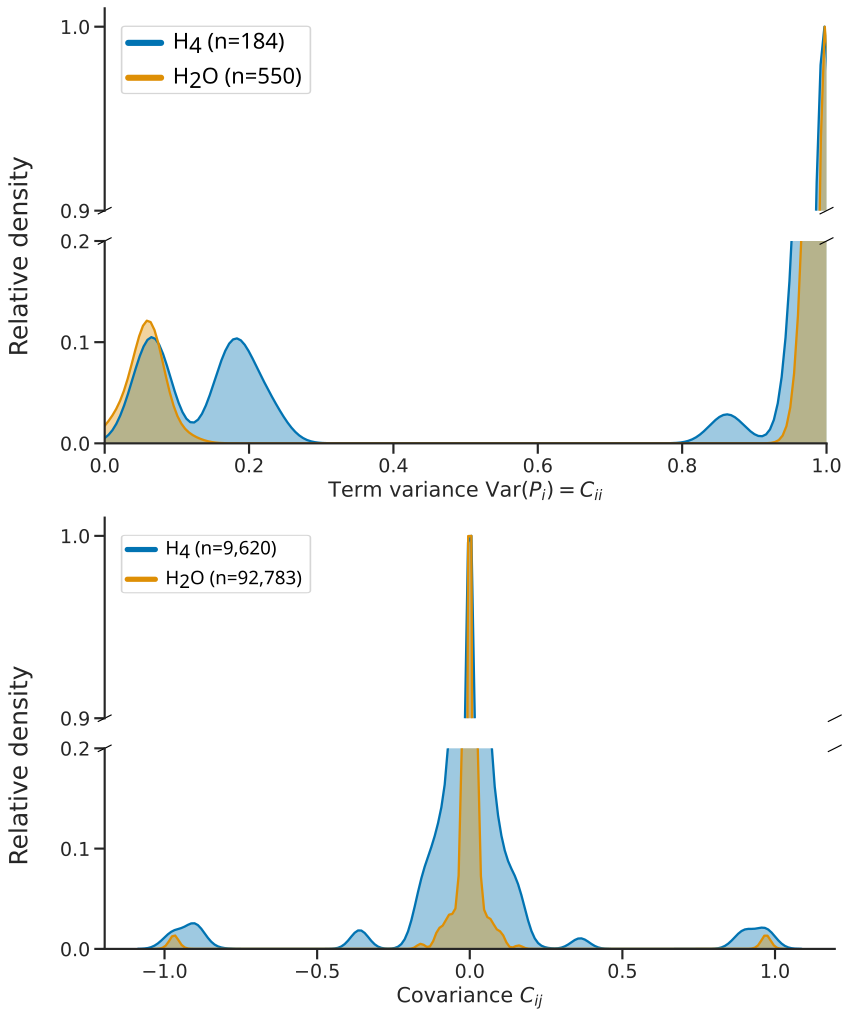}
    \caption{Distributions of the variances and covariances for the \ce{H4} and \ce{H2O} qubit Hamiltonians using the exact wavefunction; the number of data points for each distribution is shown in parentheses.
    Each density is independently scaled to a maximum of unity, allowing comparison of distribution shapes. Smoothed distributions were obtained using kernel density estimation with a Gaussian kernel and a smoothing factor of 0.8 for the covariances and of 0.2 for the variances.
    }
    \label{fig:Distro}
\end{figure}

Our analysis provides valuable insights into the QWC grouping problem and suggests promising directions for future GNN implementations. For instance, instead of coloring the complement of the commutativity graph, one could directly use the commutativity graph, which is typically sparser, and solve a clique-cover type problem. Beyond GNNs, alternative architectures such as transformers could also be explored, though they scale quadratically with the number of nodes \cite{Transformer}, in contrast to the linear edge scaling of GNN message passing. While such extensions are beyond the scope of this work, our results show that GFlowNets can still discover nontrivial solutions, such as achieving only 2\% more measurements than SI for \ce{H2O} while reducing the number of groups by 8.

\subsubsection{Effect of node ordering}\label{sec:Node}
To understand why the algorithm underperforms in generating non-overlapping groupings for \ce{H2O} in FC and the QWC instances, we analyze the SI algorithm and its approximations. SI is a greedy algorithm that maximizes the quantity $R_{SI}=\frac{M_u}{M_g}$ approximated by replacing the variances/covariances with their expectation value over the uniform spherical measure. Within this approximation, the variances of each Pauli operator are $\text{Var}(P_i)=1-\frac{1}{2^{n_q}+1}$, and the covariances, defined as $C_{ij} = \langle\psi|\hat{P}_i\hat{P}_j|\psi\rangle - \langle\psi|\hat{P}_i|\psi\rangle\langle\psi|\hat{P}_j|\psi\rangle,$ between any pair of terms is zero. Thus SI minimizes the proxy, $\hat R_{SI} = \left(\sum_\alpha\sum_i|\omega_{i,\alpha}|\right)/\left(\sum_\alpha\sqrt{\sum_i\omega_{i,\alpha}^2}\right)$. This makes SI especially efficient for molecules that approach such values and thus, especially hard to beat, as it naturally minimizes the denominator by producing an uneven distribution of groups, meaning compatible operators with larger coefficients are grouped together first, leaving terms with lower coefficients for later steps and thus reducing the square root in the denominator $\sum_\alpha\sqrt{\sum_i\omega_{i,\alpha}^2}$ with $\omega_{i,\alpha}$ the weight of the $i-$th Pauli word in group $\alpha$. In Fig. \ref{fig:Distro} we compare the distribution of variances and covariances of \ce{H4}, the molecule where GFlowNets reduces measurements the most with respect to SI, and the distribution of these values for \ce{H2O}. Fig. \ref{fig:Distro} shows the KDE using a Gaussian kernel, with bandwidths of 0.2 for the variances and 0.8 for the covariance. Each KDE is independently scaled to a maximum of unity, allowing comparison of distribution shapes; standard distributions and histograms for this comparison are available in the SM Fig. SM-5. Our findings show that precisely the distribution for \ce{H2O} corresponds to an almost ideal scenario for SI; most of the covariances between commuting Pauli operators are close to zero, while the variances are closer to unity with the exact wavefunction employed here. This underscores the difficulty of beating SI, as, in its simplicity, it becomes a great heuristic for molecular systems with variance and covariance distributions amenable to its approximations, substantially reducing measurement cost. Similar observations are found in general across the different molecules considered here in both the FC and QWC settings, as shown in Fig. SM-6. For this reason, combining heuristic ideas from SI with our GFlowNets protocol becomes an attractive approach.

{\renewcommand{\arraystretch}{1.3}
\setlength{\tabcolsep}{2pt}
\begin{table}[h!]
    \centering
    \begin{tabular}{|c||c|c||c|c|}
\hline
\begin{tabular}[c]{@{}c@{}}System [$N_P$]\\ FC\end{tabular}      & SI           & S-GFN                 & SI-ICS              & S-GFN-ICS            \\ \hline
\ce{H4} [184]                                                    & 1.015 (9)    & \textbf{0.676 (11)}   & 0.705 (9)           & \textbf{0.422 (11)}  \\ \hline
\ce{LiH} [275]                                                   & 0.276 (20)   & \textbf{0.202 (23)}   & 0.122 (20)          & \textbf{0.090 (23)}  \\ \hline
\ce{BeH2} [326]                                                  & 0.613 (18)   & \textbf{0.527 (20)}   & 0.301 (18)          & \textbf{0.252 (20)}  \\ \hline
\ce{H2O} [550]                                                   & 2.783 (30)   & \textbf{2.660 (34)}   & 1.005 (30)          & \textbf{0.854 (34)}  \\ \hline
\ce{LiH}$^*$ [630]                                                   & 0.882 (42)   & \textbf{0.602 (39)}   & 0.233 (42)          & \textbf{0.173 (39)}  \\ \hline
\ce{N2} [824]                                                    & 3.782 (26)   & \textbf{3.572 (33)}   & 1.612 (26)          & \textbf{1.558 (33)}  \\ \hline
\ce{MgO} [1616]                                                  & 8.927 (53)   & \textbf{7.670 (69)}   & \textbf{1.671 (53) }         &   1.828 (69)                \\ \hline
\ce{SiO} [1616]                                                  & 17.829 (55)  & \textbf{14.678 (64)}  & \textbf{1.992 (55)}          &   2.318 (64)                   \\ \hline \hline
\begin{tabular}[c]{@{}c@{}}System [$N_P$]\\ QWC, JW\end{tabular} & SI           & S-GFN                 & SI-ICS              & S-GFN-ICS            \\ \hline
\ce{H4} [184]                                                    & 3.993 (69)   & \textbf{3.839 (71)}   & \textbf{2.393 (69)} & 2.491 (71)           \\ \hline
\ce{LiH} [275]                                                   & 0.468 (81)   & \textbf{0.412 (74)}   & 0.283 (81)          & \textbf{0.246 (74)}  \\ \hline
\ce{BeH2} [326]                                                  & 3.977 (101)  & \textbf{3.930 (101)}  & 3.154 (101)         & \textbf{3.114 (101)} \\ \hline
\ce{H2O} [550]                                                   & 10.852 (185) & \textbf{10.599 (184)} & 7.206 (185)         & \textbf{7.021 (184)} \\ \hline

\begin{tabular}[c]{@{}c@{}}System [$N_P$]\\ QWC, BK\end{tabular} & SI           & S-GFN                 & SI-ICS              & S-GFN-ICS            \\ \hline
\ce{H4} [184]                                                    & 2.781 (56)   & \textbf{2.666 (57)}   & \textbf{1.758 (56)} & \textbf{1.758 (56)}  \\ \hline
\ce{LiH} [275]                                                   & 0.861 (75)   & \textbf{0.691 (65)}   & 0.370 (75)          & \textbf{0.356 (65)}  \\ \hline
\ce{BeH2} [326]                                                  & 2.809 (79)   & \textbf{2.760 (79)}   & \textbf{1.893 (79)} & \textbf{1.893 (79)}  \\ \hline
\ce{H2O} [550]                                                   & 22.531 (188) & \textbf{18.727 (179)} & 10.440 (188)        & \textbf{9.939 (179)} \\ \hline

\end{tabular}
    \caption{Comparison of $\varepsilon^2M(x)$ values in the FC grouping with GFlowNets protocol using the $\texttt{GINE}_w$ model for Sorted GFlowNets (S-GFN). These values are equivalent to the number of measurements (in millions) required to achieve 1 mHa accuracy. The number of groups $N_G(x)$ is reported in parentheses. Results in bold indicate the solutions with the lowest $\varepsilon^2M(x)$ for both the non-overlapping and overlapping settings. The SI-ICS overlapping method uses the groups from Sorted Insertion as a starting point, while S-GFN-ICS uses the groupings from S-GFN.
    }
    \label{tab:ExactVarS-GFN}
\end{table}}

In light of these results, we designed a variant which orders the Pauli operators in terms of the absolute value of the coefficients $|\omega_k|$, in a similar fashion to SI, which we will refer to as sorted GFlowNets (S-GFN). This variant employed the $\texttt{GINE}_w$ model and performed remarkably well owing to the covariance distributions of most of the systems studied here (see Fig. SM-6). The architectures are kept identical to those in the previous section to provide a fair comparison across different graph orderings. We include the \ce{MgO} molecule, which uses a bond distance of 1.75 {\AA}, close to the equilibrium geometry, as an extension to test this variant. For this molecule, we employ the same architecture as for \ce{SiO}, namely, \texttt{emb\_d}=2 and \texttt{hidden\_d}=24. For the molecules \ce{H4}, \ce{LiH}, \ce{BeH2} and \ce{H2O}, 5000 samples were collected; for \ce{LiH}$^*$ and \ce{N2}, 1000 samples and for \ce{MgO} and \ce{SiO}, 500 samples. For QWC groupings, we used \texttt{emb\_d}=2 and \texttt{hidden\_d}=64 for all molecules. We did not include \ce{N2}, \ce{SiO} or \ce{MgO} in the QWC search due to memory and time-scaling limitations arising from the dense conflict graphs of the QWC groupings, as we previously discussed; different implementations are required for this task.

As shown in Table \ref{tab:ExactVarS-GFN}, the non-overlapping grouping obtained with S-GFN consistently falls below the SI baseline, even for the molecules and settings where our initial formulation struggled, namely, \ce{H2O} in FC and the molecules considered in the QWC setting. The proposed variant for JW-mapped Hamiltonians in the FC setting yields an average 18.5\% reduction in measurement requirements for non-overlapping groupings compared to SI, with maximum reductions of 33\%. For overlapping groupings, we find a 12.6\% reduction when using S-GFN as the initialization compared to SI-ICS. For overlapping ICS groupings, S-GFN-ICS initialization yields up to a 40\% reduction. However, for \ce{MgO} and \ce{SiO}, the improved initialization did not yield better overlapping performance. We highlight that \ce{H4} deviates the most from the idealized scenario considered in the SI, as shown in Fig. SM-6, being the type of system where we can gain the most from our GFlowNets implementation. In the QWC setting, while the S-GFN variant outperformed the SI and SI-ICS baselines in most cases, the benefits were negligible. We attribute this behavior to the previously discussed issues with QWC, still struggling to produce a large number of valid graphs, highlighting the need for alternative approaches that directly employ the commutativity graph, which is sparser and thus scales better. Even with this modification, we observed a low percentage of valid graphs during QWC sampling, with the algorithm spending most of its time learning to produce a valid graph rather than exploring high-reward areas. Taken together, these results show that our GFlowNets implementation can produce non-overlapping groupings with lower measurement requirements than standard greedy algorithms, with potential applications towards overlapping method initialization.

\begin{figure}[h!]
    \centering
    \includegraphics[width=0.85\linewidth]{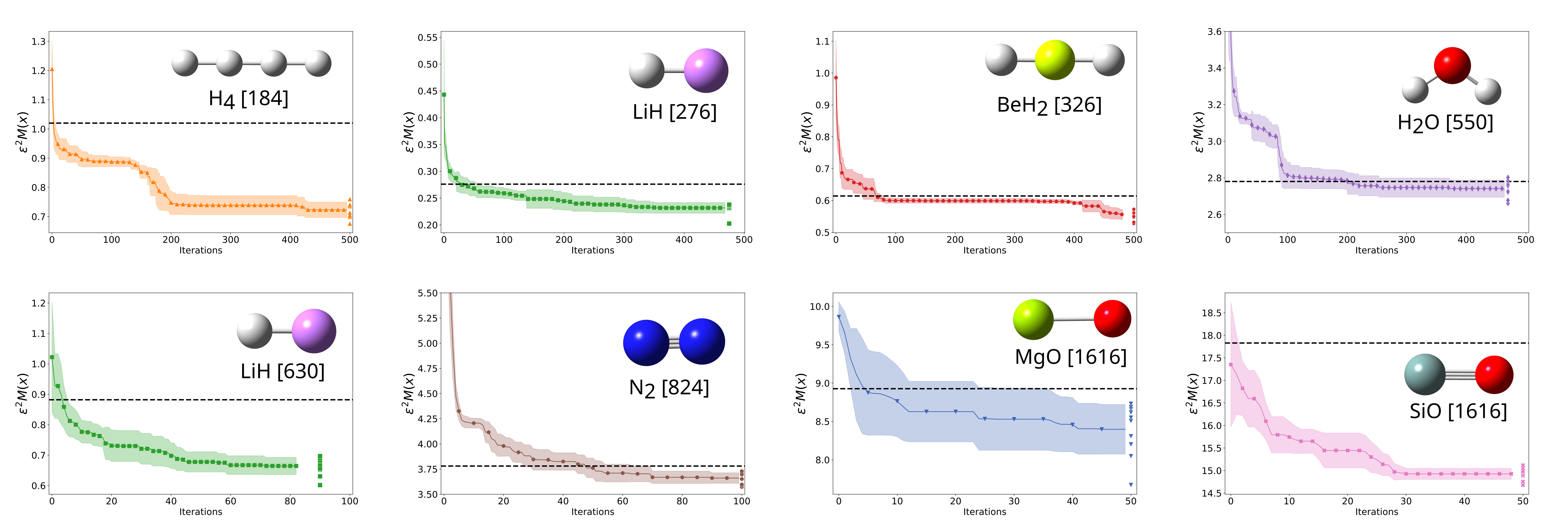}
    \caption{Average of the top 10 samples, at each iteration, according to their reward over the sampling procedure. $N_P$ is shown within the brackets. The model employed is the $\texttt{GINE}_w$ with the sorted initialization for GFlowNets (S-GFN). The points at the end of the graph are the overall best 10 samples. The Hamiltonian for LiH with $N_P=630$ was taken from Ref.~\cite{Thomson-npj}. Measurements obtained via sorted insertion are shown as a black dotted line in each graph.}
    \label{fig:CumulativeBestOrdered}
\end{figure}

To compare the behavior of the algorithm with the standard node ordering and the sorted ordering of S-GFN, we show the top 10 samples, according to their measurement value $\varepsilon^2M(x)$, throughout the iterations in Fig. \ref{fig:CumulativeBestOrdered}. Compared with Fig. \ref{fig:CumulativeBest}, we observe, for the systems in common, a clear reduction in the number of iterations required to fall below the SI threshold. For \ce{LiH}, the number of iterations to beat SI decreases from 1000 to less than 500; for \ce{BeH2} the S-GFN variant beats SI in less than 1000 iterations, compared to the standard qubit Hamiltonian ordering. Similar effects are observed in the other molecules. Remarkably, for \ce{H2O}, the S-GFN variant is able to beat the SI groupings with a very simple modification to the graph initialization, showing that the combination of the generative model with a simple heuristic can improve the efficiency of the learning algorithm. For completeness, we include the histogram of the sampled groupings and the S-GFN-ICS Pareto front results for both FC and QWC settings in Figs. SM-7 and SM-8.

\subsection{Composite rewards}\label{sec:Composite}
\subsubsection{Measurements and number of circuits}
In this section, we illustrate the impact of considering a composite reward that accounts for both the number of measurements and the number of groups. We allow $\lambda_0$, in Eq. \ref{eqn:reward}, to vary and set $\lambda_1=1$. We employ the \texttt{GINE}$_w$ model with $\texttt{emb\_d}=2$ and $\texttt{hidden\_d}=64$ architecture and collect $5,000$ samples for the \ce{H4}, \ce{LiH}, and \ce{BeH2} molecules. We limit these three Hamiltonians to the JW mapping and FC grouping. We employ the standard node ordering from the qubit Hamiltonian object in this section.

In Table \ref{tab:Custom_Rewards}, we summarize the results from this analysis, showing the grouping with the lowest measurement count, as well as the minimum number of groups found across the sampling. We consider two regimes, i) measurement-heavy when the term $\lambda_0R_M(x)$ dominates the reward, and ii) grouping-heavy when $\lambda_1R_G(x)$ does, using the lowest number of measurements and the lowest number of groups found through the measurement-only reward sampling as reference. The grouping-heavy regime, for the considered molecules, corresponds to $\sim\lambda_0 \leq (131, 57, 188)$ for \ce{H4}, \ce{LiH}, and \ce{BeH2}, respectively. The measurement-heavy regime corresponds then to $\sim\lambda_0 > (131, 57, 188)$ for each system. To complement the previous results, we also look at a grouping-only reward with $\lambda_0=0$.

For \ce{H4}, we see a clear difference in the grouping-heavy regime, with solutions containing 8 groups consistently found. When the measurement component of the reward starts to dominate, this value increases slightly. In this case, all of the solutions with the lowest measurement counts for each run are comparable, making the number of groups the main difference between them, i.e., $\varepsilon^2M(x)=0.778$ with 11 groups vs $\varepsilon^2M(x)=0.785$ with 9 groups, where constructing two fewer quantum circuits comes at a cost of less than 1\%. For \ce{LiH}, the solution with the lowest number of measurements corresponds to the most measurement-heavy reward considered with $\lambda_0=10^3$. However, as with \ce{H4}, the solutions found fall within the same range, making the number of groups the main difference in the experimental implementation. On the other extreme, we find the minimum number of groups for the grouping-only reward ($\lambda_0=0$). In the case of \ce{BeH2}, we find the minimum number of groups within the measurement-heavy regime, with 10 groups and the lowest measurement requirement with a measurement-dominated reward, beating our previous measurement-only result. When comparing these, we find that the measurement-only solution for \texttt{GINE}$_w$, gets $\varepsilon^2M(x)=0.601$ with 21 groups, while the solution with $\lambda_0=400$ and $\lambda_1=1$ gets $\varepsilon^2M(x)=0.550$ with 16 groups, showing that this reward can select a grouping with both fewer groups and a lower measurement estimate in this benchmark. In terms of the reward, this particular solution has a ratio slightly above 2:1 for $R_M(x):R_G(x)$, illustrating how the reward weights modify the sampled trade-off between measurement count and the number of groups.
%ratio of 2.346:1
{\renewcommand{\arraystretch}{1.20}
\setlength{\tabcolsep}{1pt}
\begin{table*}[]
%\resizebox{\columnwidth}{!}{
\centering
%\small
\begin{tabular}{|ccc||ccc||ccc|}
\hline
\multicolumn{3}{|c||}{\ce{H4} [184]}                                                                     & \multicolumn{3}{c||}{\ce{LiH} [275]}                                                                    & \multicolumn{3}{c|}{\ce{BeH2} [326]}                                                                   \\ \hline
\multicolumn{1}{|c|}{$\lambda_0$} & \multicolumn{1}{c|}{$\min\varepsilon^2M\ (N_G)$} & $\varepsilon^2M\ (\min{N_G})$ & \multicolumn{1}{c|}{$\lambda_0$} & \multicolumn{1}{c|}{$\min\varepsilon^2M\ (N_G)$} & $\varepsilon^2M\ (\min{N_G})$ & \multicolumn{1}{c|}{$\lambda_0$} & \multicolumn{1}{c|}{$\min\varepsilon^2M\ (N_G)$} & $\varepsilon^2M\ (\min{N_G})$ \\ \hline
\multicolumn{1}{|c|}{0}           & \multicolumn{1}{c|}{0.792 (9)}             & 1.13 (8)           & \multicolumn{1}{c|}{0}           & \multicolumn{1}{c|}{0.229 (21)}            & 0.490 (13)          & \multicolumn{1}{c|}{0}           & \multicolumn{1}{c|}{0.618 (14)}            & 0.671 (12)          \\ \hline
\multicolumn{1}{|c|}{25}          & \multicolumn{1}{c|}{0.809 (10)}            & 1.05 (8)           & \multicolumn{1}{c|}{25}          & \multicolumn{1}{c|}{0.223 (17)}            & 0.237 (16)          & \multicolumn{1}{c|}{25}          & \multicolumn{1}{c|}{0.642 (12)}            & 0.642 (12)          \\ \hline
\multicolumn{1}{|c|}{50}          & \multicolumn{1}{c|}{0.805 (11)}            & 1.92 (8)           & \multicolumn{1}{c|}{50}          & \multicolumn{1}{c|}{0.238 (18)}            & 0.418 (16)          & \multicolumn{1}{c|}{50}          & \multicolumn{1}{c|}{0.613 (17)}            & 0.800 (12)          \\ \hline
\multicolumn{1}{|c|}{75}          & \multicolumn{1}{c|}{0.795 (11)}            & 1.08 (8)           & \multicolumn{1}{c|}{75}          & \multicolumn{1}{c|}{0.238 (18)}            & 0.400 (13)          & \multicolumn{1}{c|}{75}          & \multicolumn{1}{c|}{0.638 (19)}            & 0.912 (14)          \\ \hline
\multicolumn{1}{|c|}{100}         & \multicolumn{1}{c|}{\textbf{0.778 (11)}}   & 0.785 (9)           & \multicolumn{1}{c|}{100}         & \multicolumn{1}{c|}{0.224 (20)}            & 0.236 (16)          & \multicolumn{1}{c|}{100}         & \multicolumn{1}{c|}{0.601 (16)}            & 0.686 (10)          \\ \hline
\multicolumn{1}{|c|}{200}         & \multicolumn{1}{c|}{0.786 (9)}             & 0.865 (8)           & \multicolumn{1}{c|}{200}         & \multicolumn{1}{c|}{0.229 (18)}            & 0.231 (14)          & \multicolumn{1}{c|}{200}         & \multicolumn{1}{c|}{0.631 (15)}            & 1.009 (12)          \\ \hline
\multicolumn{1}{|c|}{400}         & \multicolumn{1}{c|}{0.789 (10)}            & 0.810 (9)           & \multicolumn{1}{c|}{400}         & \multicolumn{1}{c|}{0.237 (18)}            & 0.497 (16)          & \multicolumn{1}{c|}{400}         & \multicolumn{1}{c|}{\textbf{0.550 (16)}}  &0.725 (12)          \\ \hline
\multicolumn{1}{|c|}{600}         & \multicolumn{1}{c|}{0.812 (11)}            & 0.884 (9)           & \multicolumn{1}{c|}{600}         & \multicolumn{1}{c|}{0.231 (18)}            & 0.387 (15)          & \multicolumn{1}{c|}{600}         & \multicolumn{1}{c|}{0.588 (15)}           & 0.754 (12)          \\ \hline
\multicolumn{1}{|c|}{800}         & \multicolumn{1}{c|}{0.797 (12)}            & 0.893 (9)           & \multicolumn{1}{c|}{800}         & \multicolumn{1}{c|}{0.221 (18)}            & 0.371 (13)          &  \multicolumn{1}{c|}{800}         & \multicolumn{1}{c|}{0.622 (15)}            & 0.996 (12)          \\ \hline
\multicolumn{1}{|c|}{1000}        & \multicolumn{1}{c|}{0.784 (11)}            & 0.809 (9)           & \multicolumn{1}{c|}{1000}        & \multicolumn{1}{c|}{\textbf{0.210 (18)}}            & 0.340 (15)        &  \multicolumn{1}{c|}{1000}        & \multicolumn{1}{c|}{0.581 (12)}            & 0.971 (11)          \\ \hline \hline \addlinespace

\hline
\multicolumn{1}{|c|}{\textbf{SI}}        & \multicolumn{1}{c|}{1.02 (9)}            & --           & \multicolumn{1}{c|}{\textbf{SI}}        & \multicolumn{1}{c|}{0.276 (20)}            & --        &  \multicolumn{1}{c|}{\textbf{SI}}        & \multicolumn{1}{c|}{0.614 (18)}            & --     \\\hline
\end{tabular}
\caption{Solutions with the lowest number of measurements and minimum number of groups found through the sampling for different values of $\lambda_0$ in Eq. \ref{eqn:reward}. FC grouping and JW mapping for all the Hamiltonians. In bold, we highlight the solutions with the lowest measurement count for each molecule.}
\label{tab:Custom_Rewards}
%}%end resizebox
\end{table*}}

\begin{figure}[h!]
    \centering
    \includegraphics[width=0.35\linewidth]{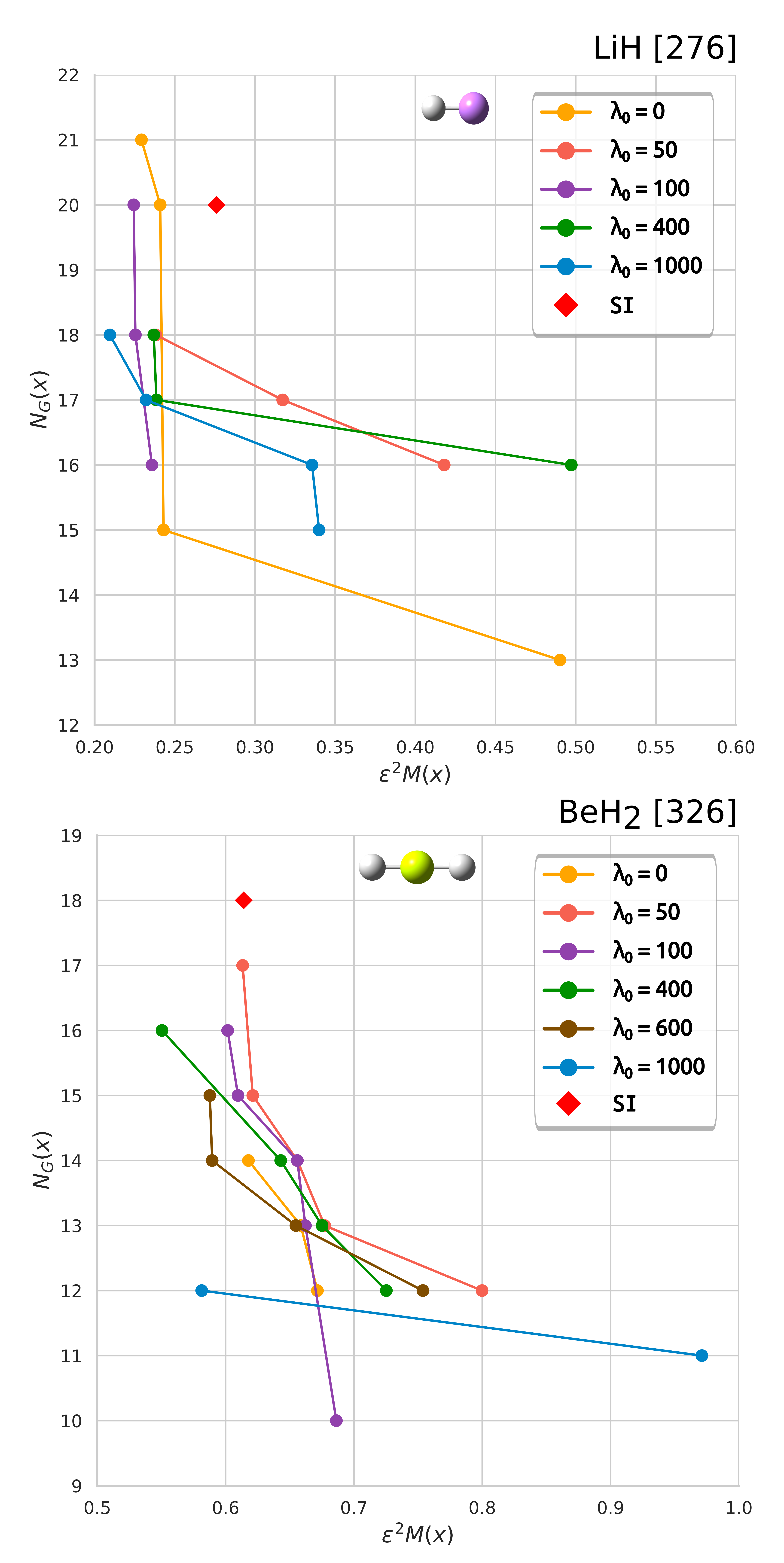}
    \caption{The Pareto fronts at different values of $\lambda_0$ on $R(x)$ (Eq. \ref{eqn:reward}) for \ce{LiH} and \ce{BeH2} with  $\lambda_1=1$ and  $\lambda_2=0$. For both systems, we used FC grouping and JW mapping.}
    \label{fig:CustRewComb}
\end{figure}

To further clarify the influence of the hyperparameters on the reward, we present in Fig. \ref{fig:CustRewComb} the Pareto fronts for selected values of $\lambda_0$ for \ce{LiH} and \ce{BeH2}. For the full histograms for each $\lambda_0$ value, see Figs. SM-9, SM-10, and SM-11  in the SM. For \ce{LiH} with $\lambda_0=0$, we observe the solutions with the minimum number of groups regardless of the measurement value, since the reward function corresponds to grouping only, making the sampler tailored for this task, although we still find competent measurement solutions. For measurement-heavy regimes, we see that the Pareto front finds better solutions in terms of measurement requirements. As for \ce{BeH2}, we see that the Pareto fronts for the measurement-heavy rewards go below $\varepsilon^2M=0.6$. Although the grouping-only reward cannot find the solution with the fewest measurements, a grouping-heavy reward that mixes measurements can, showing that multiple reward terms can change the sampled trade-off between measurement estimates and the number of groups.

\subsubsection{Measurements and number of two-qubit gates.}
Circuit costs can also be considered as their non-unit fidelity affects the overall performance of the circuit implemented on the quantum device \cite{ZachFidOverhead,quantummeasurementquantumchemistry}. To explore this direction, we consider a composite reward that includes both the number of measurements and the number of two-qubit gates across the measurement circuits for FC groups. We allow $\lambda_0$, in Eq. \ref{eqn:reward}, to vary and set $\lambda_1=0$ and $\lambda_2=10^5$. We employ the \texttt{GINE}$_w$ model with $\texttt{emb\_d}=2$ and $\texttt{hidden\_d}=64$ architecture and collect $10,000$ samples for \ce{H4} and $5000$ samples for \ce{LiH}, and \ce{BeH2}.

{\renewcommand{\arraystretch}{1.40}
\setlength{\tabcolsep}{3.5pt}
\begin{table*}[h!]
%\resizebox{\columnwidth}{!}{
\centering
%\small
\begin{tabular}{|c|cc||cc||cc|}
\hline
\multirow{3}{*}{$\lambda_0$} & \multicolumn{2}{c||}{\ce{H4} [184]}                                  & \multicolumn{2}{c||}{\ce{LiH} [275]}                                 & \multicolumn{2}{c|}{\ce{BeH2} [326]}                                \\ \cline{2-7} 
                             & \multicolumn{1}{c|}{GFlowNets}                         & GFN-ICS                          & \multicolumn{1}{c|}{GFlowNets}                         & GFN-ICS                          & \multicolumn{1}{c|}{GFlowNets}                         & GFN-ICS                          \\ \cline{2-7} 
                             & \multicolumn{1}{c|}{$\varepsilon^2M\ (N_G) / N_{2q}$} & $\varepsilon^2M\ (N_G) / N_{2q}$ & \multicolumn{1}{c|}{$\varepsilon^2M\ (N_G) / N_{2q}$} & $\varepsilon^2M\ (N_G) / N_{2q}$ & \multicolumn{1}{c|}{$\varepsilon^2M\ (N_G) / N_{2q}$} & $\varepsilon^2M\ (N_G) / N_{2q}$ \\ \hline
10                           & \multicolumn{1}{c|}{1.488 (10)/184}                 & 0.748 (10)/184                 & \multicolumn{1}{c|}{0.213 (23)/216}                 & 0.123 (23)/222                 & \multicolumn{1}{c|}{0.665 (15)/378}                 & 0.292 (15)/378                 \\ \hline
25                           & \multicolumn{1}{c|}{1.276 (8)/183}                  & 0.899 (8)/183                  & \multicolumn{1}{c|}{0.219 (22)/245}                 & 0.130 (22)/245                 & \multicolumn{1}{c|}{0.676 (15)/381}                 & 0.312 (15)/381                 \\ \hline
50                           & \multicolumn{1}{c|}{1.488 (10)/184}                 & 0.748 (10)/184                 & \multicolumn{1}{c|}{0.228 (21)/246}                 & 0.129 (21)/246                 & \multicolumn{1}{c|}{0.884 (15)/339}                 & 0.357 (15)/339                 \\ \hline
75                           & \multicolumn{1}{c|}{1.282 (9)/183}                  & 0.637 (9)/183                  & \multicolumn{1}{c|}{0.220 (24)/228}                 & 0.123 (24)/228                 & \multicolumn{1}{c|}{0.771 (15)/319}                 & 0.379 (15)/319                 \\ \hline
100                          & \multicolumn{1}{c|}{0.837 (11)/204}                 & 0.393 (11)/204                 & \multicolumn{1}{c|}{0.217 (23)/270}                 & 0.116 (23)/270                 & \multicolumn{1}{c|}{0.799 (15)/333}                 & 0.347 (15)/333                 \\ \hline
200                          & \multicolumn{1}{c|}{0.784 (10)/204}                 & 0.430 (10)/206                 & \multicolumn{1}{c|}{0.218 (24)/243}                 & 0.123 (24)/249                 & \multicolumn{1}{c|}{0.680 (17)/426}                 & 0.281 (17)/426                 \\ \hline
400                          & \multicolumn{1}{c|}{0.718 (11)/264}                 & 0.414 (11)/266                 & \multicolumn{1}{c|}{0.192 (18)/380}                 & 0.100 (18)/380                 & \multicolumn{1}{c|}{0.660 (18)/397}                 & 0.335 (18)/397                 \\ \hline
600                          & \multicolumn{1}{c|}{0.808 (10)/231}                 & 0.447 (10)/231                 & \multicolumn{1}{c|}{0.206 (18)/285}                 & 0.115 (18)/285                 & \multicolumn{1}{c|}{0.606 (14)/566}                 & 0.250 (14)/566                 \\ \hline
800                          & \multicolumn{1}{c|}{0.783 (9)/237}                  & 0.438 (9)/237                  & \multicolumn{1}{c|}{0.197 (21)/314}                 & 0.093 (21)/314                 & \multicolumn{1}{c|}{0.564 (16)/516}                 & 0.284 (16)/516                 \\ \hline
1000                         & \multicolumn{1}{c|}{0.805 (11)/246}                 & 0.407 (11)/246                 & \multicolumn{1}{c|}{0.193 (20)/353}                 & 0.099 (20)/353                 & \multicolumn{1}{c|}{0.607 (12)/424}                 & 0.284 (12)/424                 \\ \hline
 \hline \addlinespace
 \hline
 \textbf{SI}                  & \multicolumn{1}{c|}{1.02 (9)/289}                   & 0.706 (9)/289                  & \multicolumn{1}{c|}{0.276 (20)/267}                 & 0.122 (20)/267                 & \multicolumn{1}{c|}{0.614 (18)/409}                 & 0.301 (18)/409                 \\ \hline
\end{tabular}
\caption{Highest rewards solutions found through sampling for different values of $\lambda_0$ in Eq. \ref{eqn:reward} with $\lambda_2=10^5$. FC grouping and JW mapping for all the Hamiltonians. GFN-ICS uses as a starting point the highest reward solution found with GFlowNets using the \texttt{GINE}$_w$ model.}
\label{tab:Custom_Rewards2q}
%}%end resizebox
\end{table*}}

Table \ref{tab:Custom_Rewards2q} shows the $\varepsilon^2M$, $N_G$, and $N_{2q}$ values for the highest reward non-overlapping groups found through the GFlowNets sampling for \ce{H4}, \ce{LiH}, and \ce{BeH2}, alongside the GFN-ICS values resulting from using them as initialization. In this case, the measurement-heavy regime corresponds to $\sim\lambda_0 \geq (350, 100, 150)$ for \ce{H4}, \ce{LiH}, and \ce{BeH2} respectively, based on the number of two-qubit gates and measurements required for the SI method. For \ce{H4}, we see that rewards that prioritize the $R_{N_{2q}(x)}$ term lead to a reduction of more than 100 two-qubit gates for the compiled circuits. Interestingly, we find solutions for $\lambda_0=75$ that are competitive in terms of the number of measurements for the non-overlapping grouping, with 105 fewer two-qubit gates, thereby further reducing measurement costs while retaining the lower two-qubit count within the GFN-ICS approach. The solution for GFN-ICS at $\lambda_0=100$ shows a reduction of 85 two-qubit gates for the implemented circuits, while reducing measurements by 44\% with respect to SI-ICS. For \ce{LiH}, we see that as $\lambda_0$ is increased, the measurement count consistently stays below the SI baseline, at the cost of an increased number of two-qubit gates. At $\lambda_0=10$, where $R_{N_{2q}(x)}$ dominates the reward in Eq. \ref{eqn:reward}, we find a non-overlapping grouping requiring 51 fewer two-qubit gates and a lower number of measurements. Using GFN-ICS for this group increases the number of two-qubit gates from 216 to 222, still 45 gates below the SI-ICS baseline, and yields a number of measurements comparable to SI-ICS. As for \ce{BeH2}, we see a similar behavior, for rewards favoring $R_{N_{2q}(x)}$, the number of two-qubit gates is reduced with respect to the SI baselines. Importantly, after using non-overlapping groupings for GFN-ICS, we see that the reduced two-qubit gate requirements are maintained, with measurement requirements comparable to those of SI-ICS; e.g., for $\lambda_0=10$, we get in GFN-ICS $\varepsilon^2M=0.292$ with $N_{2q}=378$, and for $\lambda_0=75$ $\varepsilon^2M=0.379$ with $N_{2q}=319$ compared to $\varepsilon^2M=0.292$ and $N_{2q}=409$ with SI-ICS. These results represent a 19--37\% reduction in two-qubit gate count. Since two-qubit gates are typically the dominant contributors to circuit infidelity, this reduction represents a meaningful hardware-relevant improvement, especially given that the lower gate count is preserved after ICS initialization.

\begin{figure}[h!]
    \centering
    \includegraphics[width=\linewidth]{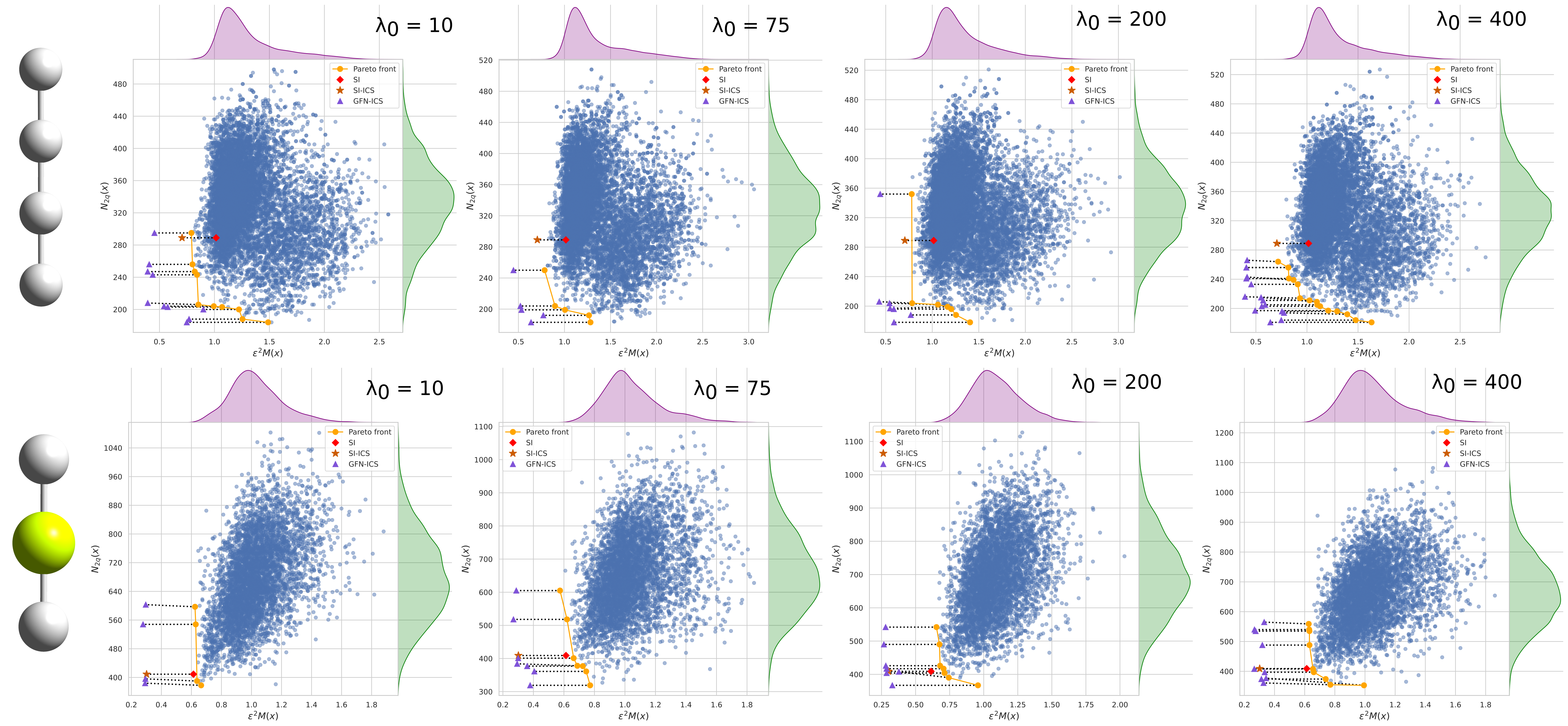}
    \caption{Histograms of the sampling process of GFlowNets with the Pareto front for different values of $\lambda_0$ in Eq. \ref{eqn:reward} with $\lambda_2=10^5$. FC grouping and JW mapping for all the Hamiltonians. 10,000 samples were used for \ce{H4} while for \ce{BeH2} 5000 were used. GFN-ICS uses as a starting point the solutions in the Pareto front found with GFlowNets using the \texttt{GINE}$_w$ model.}
    \label{fig:2QCustRewComb}
\end{figure}

To further emphasize the gains of lower two-qubit counts from non-overlapping groups generated by GFlowNets, Fig. \ref{fig:2QCustRewComb} shows the histograms for the sampling at different $\lambda_0$ values for \ce{H4} and \ce{BeH2} alongside the GFN-ICS results over the graphs in the Pareto front. We see that in \ce{H4}, several candidates have lower two-qubit counts and measurement estimates that are at or below the SI and SI-ICS values. In the case of \ce{BeH2}, we find for $\lambda_0=75$ a solution with GFN-ICS $\varepsilon^2M=0.379$ and $N_{2q}=319$, getting a 90 two-qubit gate reduction at the cost of $\sim$80,000 more measurements to reach chemical accuracy. Fig. \ref{fig:2QCustRewComb} also shows solutions for GFN-ICS better or comparable with SI-ICS in terms of measurement requirements at lower two-qubit requirements. The histograms for the additional values of $\lambda_0$ and for \ce{LiH} are presented in the SM Figs. SM-12 to SM-14.
From the samples in Fig. \ref{fig:2QCustRewComb}, we can see that different solutions can improve the SI baseline in terms of both estimated measurements and two-qubit gate count, and importantly, that the solution with the lowest count in one metric does not necessarily imply the same for the other, showing the multi-objective trade-off present in this problem. Additionally, the GFN-ICS results over the Pareto fronts suggest that the grouping with the lowest measurement estimate is not necessarily the best ICS initialization.\\

To quantify whether keeping multiple sampled groupings provides benefits towards ICS initialization beyond selecting only the lowest-measurement non-overlapping grouping found with GFlowNets, we performed a candidate-set analysis for the FC composite-reward calculations. Here, the subscript NO denotes quantities evaluated for the non-overlapping grouping before applying ICS. We first identified the single lowest-measurement non-overlapping candidate,
\begin{equation}
x_{\mathrm{NO}}^{\ast}
=
\arg\min_{x\in \mathcal{P}}
\varepsilon^2M_{\mathrm{NO}}(x),
\end{equation}
where $\mathcal{P}$ is the set of candidates appearing on the Pareto fronts defined by $\varepsilon^2M_{\mathrm{NO}}$ and $N_{2q}$. We then compared ICS initialized from $x_{\mathrm{NO}}^{\ast}$ with the best ICS result obtained from both the Pareto set $\mathcal{P}$ and an expanded candidate set $\mathcal{S}_{\mathrm{ICS}}$, constructed from the Pareto candidates together with groupings within a $25\%$ threshold in either measurement estimate or two-qubit count. We define the candidate-set gain as,
\begin{equation}
\Delta_{\mathrm{ICS}}(\mathcal{S})
=
100\%\times
\frac{
\varepsilon^2M_{\mathrm{ICS}}(x_{\mathrm{NO}}^{\ast})
-
\min_{x\in\mathcal{S}}
\varepsilon^2M_{\mathrm{ICS}}(x)
}{
\varepsilon^2M_{\mathrm{ICS}}(x_{\mathrm{NO}}^{\ast})
}.
\end{equation}
For \ce{H4}, \ce{LiH}, and \ce{BeH2}, retaining the sampled candidates from $\mathcal{S_{\mathrm{ICS}}}$ reduced the post-ICS measurement estimate by $12.5\%$, $18.9\%$, and $15.7\%$, respectively, relative to initializing ICS from $x_{\mathrm{NO}}^{\ast}$ alone. In all three cases, the best post-ICS initializer was not the grouping with the lowest non-overlapping measurement estimate, and the Spearman rank correlations between $\varepsilon^2M_{\mathrm{NO}}$ and $\varepsilon^2M_{\mathrm{ICS}}$ over $\mathcal{S}_{\mathrm{ICS}}$ were only $0.277$, $0.295$, and $0.409$ for \ce{H4}, \ce{LiH}, and \ce{BeH2}, respectively. Thus, the non-overlapping measurement estimate is only a partial predictor of downstream ICS performance. It is important to highlight that we found that retaining only the samples within the Pareto front set $\mathcal{P}$ is enough to capture most of its benefits, finding additional post-ICS measurement reductions of $12.5\%$, $15.4\%$, and $11.8\%$ for \ce{H4}, \ce{LiH}, and \ce{BeH2}, respectively, cementing the use of the Pareto front candidates as a useful strategy for grouping selection towards ICS initialization within multi-objective trade-offs. To determine whether the retained candidates correspond to genuinely distinct groupings rather than relabelings of the same partition, we used a label-invariant co-membership distance, 
\begin{equation}
D_{\mathrm{co}}(x,y)
=
\frac{1}{\binom{N_P}{2}}
\sum_{i<j}
\left|
C_x(i,j)-C_y(i,j)
\right|, \label{eqn:dist_co}
\end{equation}
where for a grouping $x$, we define
\begin{equation}
C_x(i,j)
=
\begin{cases}
1, & \text{if Pauli words } i \text{ and } j \text{ are assigned to the same group in } x,\\
0, & \text{otherwise,}
\end{cases}
\end{equation}
The denominator $\binom{N_P}{2}$, in Eq.~\ref{eqn:dist_co}, normalizes the distance by the number of unordered Pauli-word pairs, so $D_{\mathrm{co}}(x,y)$ is the fraction of pairwise co-membership relations that differ between two groupings. The near-optimal measurement subsets within 25\% showed nonzero mean co-membership distances, $D_{\mathrm{co}}=0.058$, $0.050$, and $0.026$ for \ce{H4}, \ce{LiH}, and \ce{BeH2}, respectively with maximum $D_{\mathrm{co}}$ ranging from 0.102 to 0.167, while the low-two-qubit subsets showed broader structural variation for \ce{H4} and \ce{BeH2}, with mean distances of $0.129$ and $0.112$. Even though these distances seem low, we want to emphasize that for \ce{BeH2}, a value of $D_{\mathrm{co}}=0.026$ means that of the possible $52,975$ pairs, $1,377$ of them differ in grouping assignments. Similar results are found when considering near-optimal subsets within $10\%$ for both measurements and two-qubit gate count. We also confirmed through regret analysis that the best candidate grouping from a measurement perspective differs from the two-qubit-count view, revealing trade-offs among the different rewards. For further details and additional results, we refer the reader to Sec. 10 of the SM. These results indicate that the sampled candidates have structurally distinct measurement/two-qubit trade-offs and, in these benchmarks, can improve downstream ICS initialization beyond selecting the single best non-overlapping grouping. We therefore interpret the diversity of samples, inherent to GFlowNets and possibly other generative-based models, as useful for candidate ICS initialization selection and trade-off analysis in the tested systems, rather than as a general proof that diversity alone improves measurement optimization.

\section{Conclusions}
In this study, we presented a discrete-flow-based generative model using GFlowNets for generating non-overlapping Hamiltonian groupings for quantum measurement optimization. We demonstrated that GFlowNets can sample non-overlapping grouping configurations that reduce the estimated measurement budgets relative to standard sorted-insertion heuristics \cite{SortedInsertion} across several FC benchmarks. Importantly, the non-overlapping graphs generated with GFlowNets can serve as better initializations for overlapping methods, as we show for ICS, allowing for the selection of groupings with lower measurement requirements, lower two-qubit counts, or fewer circuits to implement. Using standard node orderings, we find up to a $38\%$ reduction in measurement cost for GFN-ICS relative to SI-ICS, with an average reduction of $11\%$, on JW-mapped Hamiltonians with the algorithm struggling to beat the SI baseline for some molecular systems where the covariance distribution is centered around zero. Combining our algorithm with a simple node sorting in the S-GFN variant significantly improved the results, giving reductions of 18\% for non-overlapping groupings against SI, consistently beating the SI baseline for all molecules, regardless of FC or QWC setting and yielding post-ICS improvements of up to 40\% with an increased average reduction of 13\% for JW mapped FC groupings.
Importantly, we find that GFlowNets generates groupings with up to 37\% fewer two-qubit gate requirements, a reduction that is preserved in GFN-ICS and represents a meaningful, hardware-relevant improvement.

Beyond these empirical gains, our central conceptual contribution is a principled mechanism for sampling multiple candidate non-overlapping groupings under a specified reward instead of committing to a single heuristic output. We stress that this flexible reward selection allows the incorporation of device-specific metrics. GFlowNets operates as a sampling-based scheme, enabling the generation of multiple circuits within a single training trajectory. Unlike deterministic pipelines that yield a single partition, GFlowNets produce a diverse set of candidates that can be ranked or selected according to the specified reward function. While we demonstrate through a candidate-set analysis that solution diversity may be beneficial for ICS initialization, a more in-depth analysis of the solution space is necessary to assess whether this diversity, intrinsic to the GFlowNets algorithm, can provide a demonstrable advantage for optimization in the measurement problem and quantum resource-reduction tasks. We emphasize that our results demonstrate the usefulness of the generated samples for candidate selection and trade-off analysis, and we find that including the Pareto front set can further improve ICS initialization over the lowest-measurement non-overlapping candidate identified by GFlowNets by up to an additional $15\%$.
%We want to highlight the importance of the lower two-qubit gate counts found through our algorithm, as this improvement of up to 100 fewer two-qubit gates implies a better circuit fidelity by a factor of $F_{SI-ICS}/F_{GFN-ICS}\approx f_{2q}^{-\Delta}$ with $\Delta=N_{2q}(SI-ICS)-N_{2q}(GFN-ICS)$.

We find that our method is especially useful for quantities for which no heuristics are available, such as the total number of two-qubit gates in FC groupings. Since GFlowNets bias sampling toward high-reward regions while maintaining a distribution over terminal groupings, quantities that are not differentiable with respect to the set of actions can still be used in the GFlowNets reward, provided that they can be obtained from the terminating states. 
 Domain-specific constraints can also be seamlessly incorporated into the reward function, e.g., including the overall circuit depth, accounting for varying noise conditions, circuit fidelity overhead \cite{ZachFidOverhead}, or hardware-aware restrictions via different circuit compilation strategies\cite{SynthesisCircuitsDiffusionModel,SynthesisAlphaTensor,SynthesisQMAP,SynthesisIBMQX,SynthesisSU4,SynthesisGraphs}. In doing so, the reward can be adapted to prioritize quantities associated with a chosen compilation or hardware-cost model. Our GFlowNets protocol also becomes attractive when the studied system deviates significantly from the approximations used in the design of heuristic algorithms, such as zero covariance and unit variances used in SI. This is shown by the case of \ce{H4} and its covariance distribution analysis, which precisely reveals these deviations. For dense QWC cases, where the current model spends most of its effort learning validity, SI is the more practical default. Additional extensions to our protocol could include more flexible grouping strategies by loosening the QWC conditions. More recently, $k$-commutativity \cite{k-commutativity} and the GALIC framework \cite{Galic} were introduced to interpolate between FC and QWC measurement schemes, through $k$-commuting groupings. This method achieves roughly 20\% lower variance in energy estimates than standard grouping approaches, making it an attractive extension to our protocol. While this is outside the scope of the present paper, it can be implemented by using the $k$-commuting graph as an initial point for the GFlowNets pipeline.

While GFlowNets bypass the pre-training stage in generative models, they still require samples to learn regions with high reward. Two main limitations to scaling this class of algorithms are the trajectory length $\tau$ and the cost of evaluating the reward function. While our results for \ce{N2} are encouraging despite using fewer samples, further reducing the trajectory length will enable scaling. Furthermore, GNNs were a natural fit for the measurement problem due to the construction of the commutativity graph; however, as mentioned earlier in the paper, larger graphs require more message-passing stages, which affects the scalability of this algorithm. For a more complex reward function with multiple rewards, strategies such as those in Ref. \cite{chen2024orderpreserving} could be considered.
 Given the structure of the grouping strategy through coloring, the early stages of GFlowNets training are spent learning how to sample valid graphs before converging toward high-reward regions; see Fig.~\ref{fig:CumulativeBest}. Introducing a pre-training phase, where the model is ``warmed up'' using circuits generated by heuristic approaches such as RLF, could accelerate this process, an idea analogous to strategies previously proposed for molecular graph generation \cite{pan2024pretraining, pandey2025pretraining}. In addition, recent advances in GFlowNets \cite{takase2024gflownetfinetuningdiversecorrect, AugmentedGFlownet, liu2025nablagfn} provide promising directions to strengthen our framework. These future developments may provide mechanisms for fine-tuning, controlled sampling, and variance reduction. Leveraging precomputed covariance dictionaries from classical, efficient methods may further enhance learning efficiency and reduce the computational overhead of reward evaluation. Evaluating whether such extensions are competitive with covariance-based measurement allocation \cite{Thomson-npj} techniques remains a topic for future work.

An important point for future development is the potential to generate transferable models through our sampling strategy. Although the current protocol achieves meaningful reductions in measurement and two-qubit gate counts for FC groupings, a transferable pretrained model that can produce on-demand, high-quality, non-overlapping groupings would be of interest to the community. To reach a general, transferable model for different molecular Hamiltonians, the GFlowNets architecture would require a compatible set of actions. Meaning, a fully transferable GFlowNets model for $N_q$ Hamiltonians should be trained on the elements of the $N_q$-Pauli group, using an appropriate set of coefficients or a sufficiently diverse set of molecular Hamiltonians. Such developments would entail a great computational effort and would require GNN techniques and architectures tailored for large-scale training\cite{duancomprehensive,wang2023graph,GNNInferenceLargeScale,EfficientTrainGNNLargeGraphs,liu2024scalableGNNLargeGraphs,ScalingGNNProbabLargeScale}. While this was not the objective of the current paper, the developments from our work in conjunction with the models produced for individual molecules can serve as initializations for training such general models, considering that the associated graphs of the Hamiltonians studied here are forcefully a subgraph of the $N_q$-Pauli group.

Taken together, our results indicate that GFlowNets-based sampling is a promising direction for the initial grouping stage of quantum measurement optimization. GFlowNets, in particular, offer a compelling paradigm for complex design problems in quantum computing and beyond. They unify the strengths of generative modeling, the ability to sequentially construct solutions under domain-specific constraints, and the flexibility to sample candidates according to a user-defined reward. In the context of quantum measurement optimization, this combination can identify candidate groupings with lower estimated costs in the benchmarks studied here and expose reward-dependent trade-offs among viable groupings, e.g., the solution with the lowest two-qubit count is not the one with the lowest measurement estimates. This problem, on its own, provides a fertile testing ground for advancing the GFlowNets methodology itself, motivating the design of more efficient training protocols, richer reward structures, and scalable implementations tailored to future hardware.
We hope that this work encourages further efforts to integrate flow-based generative models into optimization tasks across quantum computing. 

% -------------- -------------- -------------- -------------- --------------
\section*{Acknowledgments}
This research was enabled by support from the Digital Research Alliance of Canada and NSERC Discovery Grant No. RGPIN-2024-06594. This research was partly enabled by Compute Ontario (computeontario.ca) and the Digital Research Alliance of Canada (alliancecan.ca) support. 

\section*{Data availability}
The code for our GFlowNets implementation is available at \href{https://github.com/ChemAI-Lab/GFlowNets-MOpt}{https://github.com/ChemAI-Lab/GFlowNets-MOpt}. This repository contains the necessary code to reproduce the reported results used in this work.
We used Torch-geometric (2.7.0),  PySCF (2.12.1), NetworkX (3.6.1), Tequila (1.9.9), and OpenFermion (1.7.1). For further details, please consult the \texttt{requirements.txt} file in the GitHub repository and the SM.
For convenience, the sampled graphs and outputs for the FC groupings obtained with the \texttt{GINE}$_w$ model in Table~\ref{tab:ExactVarFC}, as well as the sampled graphs, outputs, and models for the measurement plus two-qubit gate count composite-reward results in Table~\ref{tab:Custom_Rewards2q}, are available through Zenodo at \href{https://doi.org/10.5281/zenodo.21050287}{doi.org/10.5281/zenodo.21050287}.

%%%%%%%%%%%%%%%%%%%%%%%%%%%%%%%%%%%%%%%%%%%%%%%%%%%%%%%%%%%%%%%%%%%%%
%% If you are using classical BibTeX rather than biblatex,
%% remove the \printbibliography and uncomment the \bibliograpy one
%%%%%%%%%%%%%%%%%%%%%%%%%%%%%%%%%%%%%%%%%%%%%%%%%%%%%%%%%%%%%%%%%%%%%
%\printbibliography
\bibliography{references.bib}
% \section*{TOC Graphics}
%     \centering
%     \includegraphics[width=8.25cm]{Figures/TOC.png}\\
%     For Table of Contents only.

% \newpage
% \includepdf[pages=-]{SM.pdf}

\end{document}